\documentclass[10pt,letterpaper]{article}
\usepackage[top=0.85in,left=0.65in,footskip=0.85in,marginparwidth=2in]{geometry}
\usepackage[utf8]{inputenc}
\usepackage{float}
\usepackage{enumitem}
\usepackage{booktabs}
\usepackage{url}
\usepackage{amsmath}
\usepackage{bbm}
\usepackage{subcaption}
\usepackage{amssymb}
\newcommand*{\bv}[1]{\ensuremath \boldsymbol{#1}}
\newcommand*{\thb}{\bv{\theta}}%
\newcommand*{\phb}{\bv{\phi}}%
\newcommand*{\cov}{\bv{\kappa}}
\newcommand*{\Cov}{\mathbf{\Sigma}}
\newcommand*{\icone}{\bv{C_{pre}}^1}
\newcommand*{\icN}{\bv{C_{pre}}^T}
\newcommand*{\ivone}{\bv{C_{post}}^1}
\newcommand*{\ivM}{\bv{C_{post}}^{T^*}}

\newcommand*{\ipre}{\bv{C_{pre}}}
\newcommand*{\ipost}{\bv{C_{post}}}
\newcommand*{\Wpre}{\bv{W_{pre}}}
\newcommand*{\Wpost}{\bv{W_{post}}}
\newcommand*{\Wone}{\bv{W_{pre}}^1}
\newcommand*{\WN}{\bv{W_{pre}}^T}
\newcommand*{\Wvone}{\bv{W_{post}}^1}
\newcommand*{\WvM}{\bv{W_{post}}^{T^*}}

\usepackage{cite}
\usepackage{nameref,hyperref}
\usepackage{microtype}
\DisableLigatures[f]{encoding = *, family = * }
\raggedright
\usepackage{makecell}
\usepackage{changepage}
\usepackage[aboveskip=1pt,labelfont=bf,labelsep=period,singlelinecheck=off]{caption}
\makeatletter
\renewcommand{\@biblabel}[1]{\quad#1.}
\makeatother

\usepackage{lastpage,fancyhdr,graphicx}
\usepackage{epstopdf}
\fancyheadoffset[L]{0.in}
\fancyfootoffset[L]{0.in}

\usepackage{color}
\definecolor{Gray}{gray}{.25}
\usepackage{color, colortbl}
\definecolor{blue}{rgb}{0.639,0.784,0.906}
\usepackage{graphicx}
\usepackage{sidecap}
\usepackage{wrapfig}
\usepackage[pscoord]{eso-pic}
\usepackage[fulladjust]{marginnote}
\reversemarginpar

\begin{document}

\begin{flushleft}
{\Large
\textbf\newline{Shut and re-open: the role of schools in the spread of COVID-19 in Europe}
}
\newline
\\
Helena B. Stage\textsuperscript{1,2,$\dagger$,*},
Joseph Shingleton\textsuperscript{3,$\dagger$,*},
Sanmitra Ghosh\textsuperscript{4,$\dagger$},
Francesca Scarabel\textsuperscript{1,2,5,6},
Lorenzo Pellis\textsuperscript{1,2,7},
Thomas Finnie\textsuperscript{3}
\\
\bigskip
\bf{1} Department of Mathematics, The University of Manchester, UK
\\
\bf{2} Joint UNIversities Pandemic and Epidemiological Research, \url{https://maths.org/juniper/}
\\
\bf{3} Emergency Response Department, Public Health England, UK
\\
\bf{4} MRC Biostatistics Unit, University of Cambridge, Cambridge, UK
\\
\bf{5} Laboratory of Industrial and Applied Mathematics, Department of Mathematics and Statistics, York University, Toronto Ontario, Canada
\\
\bf{6} CDLab - Computational Dynamics Laboratory, Department of Mathematics, Computer Science and Physics, University of Udine, Italy
\\
\bf{7} The Alan Turing Institute, London, UK

\bigskip
* joseph.shingleton@phe.gov.uk, helena.stage@manchester.ac.uk\\
$\dagger$ These authors contributed equally to the production of this manuscript.
\end{flushleft}

\section*{Abstract}
We investigate the effect of school closure and subsequent reopening on the transmission of COVID-19, by considering Denmark, Norway, Sweden, and German states as case studies. By comparing the growth rates in daily hospitalisations or confirmed cases under different interventions, we provide evidence that school closures contribute to a reduction in the growth rate approximately 7 days after implementation. Limited school attendance, such as older students sitting exams or the partial return of younger year groups, does not appear to significantly affect community transmission. In countries where community transmission is generally low, such as Denmark or Norway, a large-scale reopening of schools while controlling or suppressing the epidemic appears feasible. However, school reopening can contribute to statistically significant increases in the growth rate in countries like Germany, where community transmission is relatively high. In all regions, a combination of low classroom occupancy and robust test-and-trace measures were in place. Our findings underscore the need for a cautious evaluation of reopening strategies.

\paragraph{Keywords:}COVID-19, school closure, school reopening, non-pharmaceutical interventions

\section*{Introduction}
Throughout the course of the ongoing COVID-19 pandemic, the role of young people and children in transmission has been a question of particular concern \cite{Lee2020, Ludvigsson2020}. This question is not only motivated by the goal of protecting the younger generations; it is also known from other respiratory diseases that, because younger people tend to have more prolonged and physical contacts among themselves \cite{Mossong2008}, they pose a greater risk of infection to each other as well as being likely to introduce the infection to their respective households and so can drive the epidemic \cite{House2011, Cauchemez2008}. Consequently, school closure is often one of the first measures considered when non-pharmaceutical interventions are needed \cite{Viner2020}. However, during the COVID-19 pandemic it has often been implemented concurrently with other measures, making it difficult to assess its individual impact \cite{Flaxman2020, Chen2020}.\newline
Many of the challenges inherent in quantifying the impact of closure remain when policy-makers subsequently turn to the reopening of schools. Reopening presents a myriad of further questions, such as the ages of those returning, the physical circumstances and timing of their return, and the necessary conditions which must be met on a community level before a return can be deemed safe enough. For new or emerging infections, answers to these questions require new efforts to establish an age-stratified understanding of the infection and transmission dynamics \cite{Davies2020, Gudbjartsson2020}.\newline

Earlier work exists concerning the effectiveness of school closure as a means of controlling the spread of COVID-19, with mixed conclusions depending on the studied age group, country, and modelling assumptions \cite{Flaxman2020, Chen2020, Viner2020, Zhang2020a, Brauner2020, Banholzer2020, Davies2020}. These sought to estimate the impact of school closure on nationwide transmission levels, be it for instance a reduction in the peak number of cases or the timing thereof. The challenges and questions related to school reopening have also been addressed from a theoretical or modelling perspective of scenario planning \cite{Domenico2020, Keeling2020, Panovska-Griffiths2020}, estimating the number of new or severely ill cases resulting from school reopening. These use varying assumptions of the underlying community transmission, and consider various scenarios of the ages and timing of students returning to school.\newline
While such models are valuable means of quantifying the expected impact of measures without increasing the risk of exposing the wider community to infection, their inclusion of other interventions is necessarily limited. The times surrounding school closure and reopening have seen a myriad of other non-pharmaceutical interventions being implemented. Through social contact patterns, observed changes in community transmission are the result of non-trivial, underlying interactions between current interventions. We believe our work fills an important knowledge gap in the literature by addressing the context of school interventions alongside other measures, and the \textit{de facto} impact of schools in a broader framework of non-pharmaceutical interventions.\newline
School closure and reopening not only affect transmission occurring on school premises; they also affect (and are affected by) community transmission, transmission within households with young children, and wider measures taken to monitor and curb an outbreak. In addition to the well-being of children, school interventions also impact a nation's workforce via the time dedicated to childcare. It must be remembered that the observed effects of these interventions are a product of underlying testing, reporting, and isolation (or other physical or social distancing) measures.\newline

The aim of this work is to carry out a comparative analysis of school interventions, making use of the diversity of available data streams, to serve as a complement to theoretical modelling efforts. Ours is a data-driven approach which seeks not to establish the individual role of schools interventions on outbreak management, but instead assesses their impact in the context of wider societal interventions.
Specifically, we wish to examine roles in transmission played by a) the different age cohorts of students, b) the timing of the school interventions (closure and reopening), and c) the background or community incidence. We hope these results can serve as a series of lessons learned from nations which have already reopened schools.\newline
For school closure, we address these questions by estimating the time between intervention and a response being observed in the recorded data, as well as the changes in the growth rate pre- and post-intervention. For school reopening, we track the growth rate in cases over the intervention timeline and search for correlations between these interventions and changes in the growth rate.

\section*{Methods}

\subsection*{Data selection criteria}
We consider four countries: Denmark, Germany, Norway, and Sweden due to their geographic proximity, demographic similarities, and the relative timing and scope of their interventions to allow for a better comparison. We distinguish between countries with medium-to-high (Germany and Sweden) and low (Denmark and Norway) levels of community transmission on the basis of daily COVID-19 cases, rather than cases per capita \cite{ecdc}. This is motivated by the feasibility of testing, tracing, and isolating cases, which need not scale with population size. At the time of school closure, we saw the following cumulative cases: 800 (Denmark), 6500 (Germany), 1100 (Norway), and 1400 (Sweden).
By the time schools started reopening, the total cases had risen to: 7000 (Denmark), 158000 (Germany), 7200 (Norway), and 19400 (Sweden on the day German schools first reopened). Only German states with at least 50 cases at the point of school closure, and at least 10 days of non-zero daily cases prior to closures, have been selected for analysis.\newline
We considered hospital admissions as the primary data source in our analyses, where the numbers were available and sufficiently large to do so. All studied countries expanded their hospital surge capacities to accommodate patients to a sufficient degree that we are not aware of instances of COVID-19 patients being turned away. Since clinically ill patients are unlikely not to present themselves for treatment at hospital, admissions data are a practical measure of community infection and, unlike confirmed cases, are not as susceptible to variable testing rates in the wider population. However, by studying only a subset of the entire population, the data will be biased toward older and sicker individuals which may for example lengthen the delay from an intervention to a visible signal in the data.\newline
Confirmed cases were used in situations where hospitalisation data were not available or insufficient to reliably infer the effect of interventions - this was particularly relevant in the case of school reopening, which has predominantly been recommended in communities with significantly reduced daily incidence (and hence hospitalisation) counts. Although we do not correct for variable testing rates in the confirmed cases, we have sought out data sets where there was evidence of consistent and thorough testing. However, we acknowledge this was a challenge for most countries in early March. We document the number of tests carried out and comment further on the reliability of confirmed cases as representative of the community epidemic in the Supplementary Material (S3, Testing data).\newline
Consistent test numbers for both Germany and Norway around the time of school reopening (see Supplementary Material, S3) suggest that the confirmed number of cases is less prone to biases than earlier in the pandemic. We are therefore less concerned about using these data streams in a reopening context.\newline

\subsection*{National data streams}
The effect of school closure was estimated using hospitalisation data for Denmark and Norway, and daily confirmed cases for Germany and Sweden. Given other interventions were implemented after school closure and their effect is hard to disentangle, it is implicitly assumed that the inversion of epidemic trend from growing to declining is not solely a result of school closure. Therefore, the analysis of the impact of school closure is restricted to data before the peak in reported cases or hospital admissions.\newline
Denmark reopened schools quickly enough following sweeping nationwide interventions that hospitalisation data could still be used, though we cross-checked these findings by analysing confirmed cases. Official estimates at the time suggest a delay from infection to hospitalisation of 10 to 14 days \cite{StatensSerumInstitut2020}.
Official Norwegian estimates suggest this same delay to be 14 days \cite{NOmodel}. As Norwegian hospitalisation data were too sparse to reliably infer the effect of school reopening, daily confirmed cases were analysed instead.\newline
In Germany, daily confirmed cases are reported specifically for students under 18 in schools, kindergartens, holiday camps, after school clubs, etc.\ as well as for the staff working in these facilities. We used these numbers, and national hospital admissions, to analyse school reopening instead of population aggregates on the state or federal level.\newline 
References for each data stream, together with further discussion on their limitations, are provided in the Supplementary Material (S1, Data availability).

\subsection*{Estimating the effect of school closure by simulating the early epidemic}

Our aim is to assess the impact of school closure on transmission dynamics. This includes any associated changes in behaviour (e.g.\ parents not accompanying children to school or working from home due to caring responsibilities), as well as all other interventions (if any) occurring on the same day, which cannot be disentangled from school closure. The impact is assessed by comparing differences in the growth rate of cases or hospitalisations before and after the intervention.\newline
Rather than na\"{i}vely comparing the growth rate at different points in time, we follow a more sophisticated procedure with the aim of separating the decrease in growth rate due to interventions implemented before school closure, and the impact of school closure itself. 
This is achieved by generating a \textit{counterfactual} projection of daily cases or hospital admissions, which accounts inasmuch as possible for events prior to, but excluding, school closure, and identifying when there is a clear deviation between the real data and such a projection. We then compare the growth rate observed in the real data and in the modelled counterfactual at the time when this deviation is detected, and interpret this difference as the likely impact of school closure.\newline

To construct the counterfactual, we use a compartmental ODE model fitted to the pre-intervention data. However, fitting a model to data from the earlier part of the epidemic is extremely challenging since the observations are generally scarce, noisy and coloured through various reporting issues, in particular systematic ones, such as a strong weekend effect.\newline
Fitting a simple compartmental model without accounting for these factors will result in parameter estimates that are systematically biased \cite{Brynjarsdottir2014}. These inaccuracies in parameter estimates propagate to any projection drawn from the model. Mitigating this model discrepancy in the fitting process is an area of active research; see \cite{Lei2020} for a recent review. Generating a counterfactual projection using a compartmental model, without compensating for such discrepancies, will erroneously understate or overstate the effect of the intervention.

To alleviate the challenges brought on by scarce and noisy data we argue that, given the 4.8 day mean incubation period for SARS-CoV-2 \cite{Pellis2020a}, we expect the impact of any intervention to be delayed by at least 5 days, and in particular we expect cases on the first 5 days following school closure to predominantly reflect only earlier interventions (whether imposed or not, e.g.\ spontaneous physical distancing). We then use a two-step approach for fitting an ODE model and correcting for the discrepancy between model and data as follows. In the first step, a selection of sample trajectories are generated via Approximate Bayesian Computation (ABC) fitting of the ODE from the first day of data until 5 days after school closure. In a second post-processing step, a Bayesian regression model is then trained on the same data used to fit the ODE model, while using the sampled trajectories as the covariates (inputs). Essentially, the regression model attempts to capture the structural part of the discrepancy between ODE simulations and the observed data (predominantly, potential deviation from exponential growth and weekend effects). We formulate this regression model as a Gaussian process (GP) with a Negative Binomial likelihood. Once trained, the regression model is used to project the trajectory of cases for the time period following the 5 days after school closure. This projection is then treated as the desired counterfactual.

We identify the first day on which there is a clear and sustained deviation from the modelled data, hereafter referred to as the \textit{response date}. Such a deviation must (a) occur more than 5 days after the date of school closure, (b) fall outside of the 75\textsuperscript{th} percentile of the projected data, and (c) persist in doing so for at least 5 days. The time window from school closure to response date defines the \textit{lag time} (Table \ref{tab1}, column 2), which runs from the date of closure (acting as the zeroth day) up to but not including the response date.\newline
The growth rates are obtained as point estimates (see following description of the instantaneous growth rate) at the time of school closure, for the observed data ($r_{pre}^{obs}$, Table \ref{tab1}, column 3), and at the response date for both the modelled ($r_{post}^{mod}$, Table \ref{tab1}, column 4) and the observed data ($r_{post}^{obs}$, Table \ref{tab1}, column 5).
The relative changes in the estimated growth rates between modelled and observed data at the response date can be used to assess the impact of school closure. The observed growth rate at the time of school closure can be used to cross-check the growth rate in the modelled data; these could be significantly different if the impact of interventions prior to school closure had a strong effect on transmission which the GP was able to capture in the counterfactual projection. However, a causal link cannot be established between the interventions and the growth rates, calling for a cautious interpretation of the specific values of the growth rates and the reductions therein.\newline

The ABC fitting of the SEIR model was carried out using the PyGOM package for Python \cite{pygom}. The GP regression method, devised as a Bayesian latent variable approach, was carried out using the PyMC3 probabilistic programming package for Python \cite{Salvatier2016}. Further details about the introduced methods can be found in the Supplementary Material (S2, Numerical methods).

\subsection*{Estimating the effect of closure and reopening using the instantaneous growth rate}
With the number of sequential changes in interventions and loosened restrictions on personal movement and the operation of businesses, it is misleading to estimate a constant growth rate in new cases before and after schools reopened. We therefore consider a method whereby the growth rate can be quantified following successive changed measures. A smoother $\rho(t)$ is applied to the data over time $t$, such that the instantaneous growth rate is $\rho^\prime(t)$ (cf.\ a constant value in a phase of pure exponential growth or decline). It is assumed that the daily new confirmed cases (or daily new hospital admissions) $c(t)$ grow or decay exponentially, with noise added to account for small case numbers, i.e. $c(t) \propto e^{\rho(t)}$. To estimate $\rho^\prime(t)$ we adapt a General Additive Model (GAM) from the R package \textit{mgcv}, using a Negative Binomial family with canonical link \cite{mgcv}. Smoothing is achieved using default thin plate regression splines.\newline
Where case numbers are sufficiently high, and there is a clear weekend effect in the reporting of data, a weekend effect has been accounted for in the GAM by the addition of a fixed effect on those days of the week. Specifically, we add a quantity $\omega_d$ for each day of the week $d\in [1,7]$, such that cases follow $c(t) \propto e^{\omega_d+\rho(t)}$. Setting $\omega_d=0$ for all but two days yields a weekend-specific description, and $\omega_d=0\ \forall\ d$ recovers the GAM with no day-of-week effects. This method has previously been used in \cite{Pellis2020a}.\newline

In the case of school closure, the GAM approach with a weekend effect has been used to estimate the instantaneous growth rate at different points in time.
In the case of school reopening, the instantaneous growth rate with a day-of-week effect has been used to identify trends in the data. Our aim here is to assess if there is any correlation between changes in the growth rate, and the timing of school reopening.

\section*{Results}
As many of our findings are based on the premise of analysing interventions at different points in time, or in different geographical regions, all results are inherently conditional on the assumptions of \textit{stability} and \textit{homogeneity}. Firstly, in order to make comparisons throughout the same time series, we assume that the only changes to behaviour are those directed by changing public guidelines, and that adherence to these is constant throughout (stability). Secondly, in order to compare different regions of the same country, we must assume that there are no fundamental differences in adherence, testing, implementation of national policies, or similar such aspects (homogeneity). Deviations from these assumptions are taken to be too small to affect the data in a way to qualitatively alter our conclusions.

\subsection*{Closing of schools in Germany}
We consider the date of school closure as the first day on which all schools in a state were closed as a response to state or national government intervention. In most cases, however, there were local school closures prior to enforced closures. Furthermore, most primary schools continued to be open to both vulnerable children and the children of key workers after national and state closures.\newline
Table 1 provides an overview of the observed changes in the daily growth rates in the period during and after school closures. These growth rates are consistent with previous estimates \cite{Dehning2020}.

\newcommand{\ra}[1]{\renewcommand{\arraystretch}{#1}}

\begin{table}[H]
\ra{1.3}
\centering
\begin{tabular}{@{}lccccc@{}}
\toprule
\textbf{State} & \makecell{Lag time \\ (days)} & \makecell{$r_{pre}^{obs}$ (day$^{-1}$)} &  \makecell{$r_{post}^{mod}$ (day$^{-1}$)} & \makecell{$r_{post}^{obs}$ (day$^{-1}$)} & \makecell{$1-r_{post}^{obs}/r_{post}^{mod}$}\\
\cmidrule{1-6}
Baden-W{\"u}rttemberg & 8 &\makecell{0.143 \\(0.104 - 0.182)}& \makecell{0.167 \\(0.148 - 0.185)}& \makecell{0.051 \\(0.013 - 0.089)} & \makecell{69\%\\
(40 - 93)}\\ 
Bavaria & 8 &\makecell{0.216 \\(0.176 - 0.255)}& \makecell{0.214 \\(0.208 - 0.221)}& \makecell{0.109 \\(0.072 - 0.146)} & \makecell{49\%\\
(29 - 67)}\\ 
Berlin & --$*$ &\makecell{0.145 \\(0.103 - 0.187)}& -- & -- & --\\ 
Hesse & 7 &\makecell{0.251 \\(0.195 - 0.308)}& \makecell{0.274 \\(0.265 - 0.283)}& \makecell{0.067 \\(0.017 - 0.117)} & \makecell{75\%\\
(56 - 94)}\\ 
Lower Saxony & 7 &\makecell{0.223 \\(0.179 - 0.267)}& \makecell{0.229 \\(0.213 - 0.244)}& \makecell{0.069 \\(0.032 - 0.107)} & \makecell{70\%\\
(50 - 87)}\\ 
North Rhine-Westphalia & 6 &\makecell{0.192 \\(0.156 - 0.228)}& \makecell{0.206 \\(0.200 - 0.213)}& \makecell{0.061\\(0.026 - 0.096)} & \makecell{70\%\\
(52 - 88)}\\ 
Rhineland-Palatinate& 7 &\makecell{0.257 \\(0.205 - 0.310)}& \makecell{0.235 \\(0.211 - 0.259)}& \makecell{0.043 \\(0.001 - 0.086)} & \makecell{82\%\\
(59 - 100)}\\
\bottomrule
\end{tabular}
\caption{\color{Gray}
Estimated lag time and pre- and post-intervention (and for the latter, modelled and observed) daily growth rates in different German states, and relative change between the modelled and observed growth rate. The 95\% credible intervals (CrI) are given in brackets. Their equivalent formulation as doubling times can be found in the Supplementary Material (Table S3). Sensitivity analysis of the training period on the lag time suggests these can vary by up to two days (see Supplementary Material, S2.7). The overlapping CrIs between the pre-intervention and the post-intervention modelled growth rates suggest a limited deviation from exponential growth between the day of school closure and the end of the training window.\newline
$^*$The peak in daily incidence is reached before a response is seen in the data. A lag time which may be attributable to school closures therefore cannot be determined.}
\label{tab1}
\end{table}

All states in Germany saw a reduction in growth rate after the closure of schools, typically after a delay of 7 days, or about 1.2 generations \cite{Ferretti2020}. With the exception of Baden-W\"{u}rttemberg and Berlin, all German states closed schools on March 16\textsuperscript{th}. As this was a Monday, we set the effective date of school closures as Saturday March 14\textsuperscript{th}, under the assumption that school activity is significantly reduced on weekends. Schools in Baden-W\"{u}rttemberg and Berlin closed on Tuesday March 17\textsuperscript{th}.
It should be noted that all states experienced further interventions around the same time as school closures. The presence of concurrent interventions makes it difficult to attribute the fall in cases solely to the closure of schools, and it is likely that a combination of factors contributed to the observed decay in growth rate. However, comparison between Baden-W\"{u}rttemberg and North Rhine-Westphalia, which saw similar case numbers, yields comparable lags and overall trajectories of the epidemic curves when accounting for the three-day delay in school closures in Baden-W\"urttenberg (see Supplementary Material, Figure S9). This is indicative of school closures being at least partially responsible for the reduction in growth rate.\newline

The reduction between the modelled and observed post-response growth rates serves as a measure of the overall effectiveness of interventions (Table \ref{tab1}, column 6). Overall, lower relative reductions in the growth rate are weakly correlated with states which had higher (daily and cumulative) incidence counts at the time of intervention (Baden-W\"{u}rttemberg, Bavaria, and North Rhine-Westphalia). This supports the generally held expectation that non-pharmaceutical interventions are more effective when implemented early, capable of breaking transmission chains while community transmission is relatively low.\newline

The states of Hesse and Rhineland-Palatinate allowed students aged 18-19 to sit in-school examinations in late March, under strict social distancing measures and other precautions.
Neither of the states permitting examinations saw any less evident reduction in growth rates compared to states which had similar case numbers prior to school closure, but where exams did not take place during this time period (e.g.\ Lower Saxony). 
Further, the largest reduction in the growth rate was observed in Rhineland-Palatinate. Assuming stability and homogeneity, this suggests that under controlled conditions with limited social interaction, the beginning of the examination period for older students was likely not a significant driver of epidemic growth. We cannot comment on the full effect of the entire examination period.
We include the detailed results from the highlighted German states in Figure \ref{main_closing} and a timeline of key interventions from \cite{WHONPI} below, with the remaining states detailed in the Supplementary Material (S4, School closures analyses).
\begin{itemize}
    \item 10/03 - Banned gatherings of more than 1000 people (DE-G1).
    \item 14/03 - \textbf{Hesse, Lower Saxony, and Rhineland-Palatinate closed schools} (effective date, DE-S1).
    \item 16/03 - Shut borders with France (FR), Switzerland (CH), Austria (AT), Denmark (DK) and Luxembourg (LU) (DE-B1); closure of non-essential business and public service (DE-P1).
    \item 17/03 - \textbf{Baden-W{\"u}rttemberg closed schools} (DE-S2); shut borders with EU (DE-B2).
    \item 22/03 - National stay at home orders, with exceptions for essential trips, and banned gatherings of more than 2 people (DE-G2); banned all social events and gatherings (DE-P2); closure of non-essential retail and leisure, with exceptions for restaurant takeout (DE-R1).
\end{itemize}

\begin{figure}[H]
\centering
\includegraphics[width=170mm]{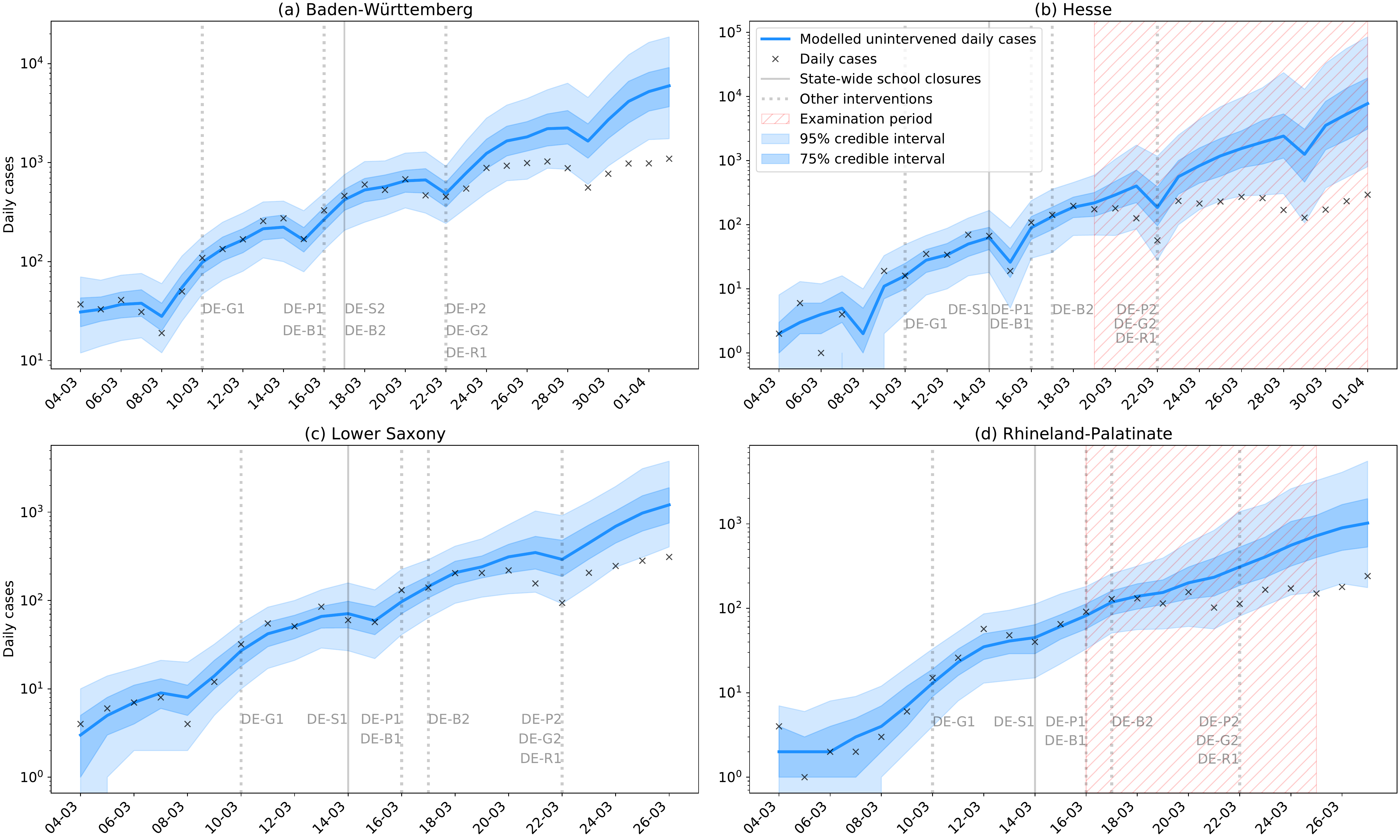}
\caption{\color{Gray}Modelled and observed cases in (a) Baden-W{\"u}rttemberg, (b) Hesse, (c) Lower Saxony, and (d) Rhineland-Palatinate. Hesse and Rhineland-Palatinate (Figure \ref{main_closing} (b), (d)), where final year high school exams took place in late March, saw a similar response to interventions as other German states with moderate incidence (Figure \ref{main_closing} (c)) where exams did not take place at that time. While there is insufficient scope in the data to assess the effect of the full examination period, we should in principle be able to detect a signal related to the beginning of the examination period. Assuming stability and homogeneity across states, and given the lack of such a signal, it is unlikely that these exams significantly contributed to the overall outbreak.} 
\label{main_closing}
\end{figure}

\subsection*{Closing of schools in Denmark, Norway, and Sweden}
 
In all three countries there were provisions in place to allow key workers’ children to continue attending school.
Hospital admissions were analysed for Denmark and Norway, as testing was deemed too variable during this time period (see Supplementary Material, S3) to reliably use confirmed cases. However, the expected lag time from infection to hospital admission in Denmark and Norway is 10-14 days \cite{StatensSerumInstitut2020, NOmodel}, whereby any signal observed in the data is too early to be attributable to school closures.
For completeness, we include the fits to daily hospital admissions in the Supplementary Material (S4, School closure analyses).\newline

\begin{figure}[h]
\centering
\includegraphics[width=130mm]{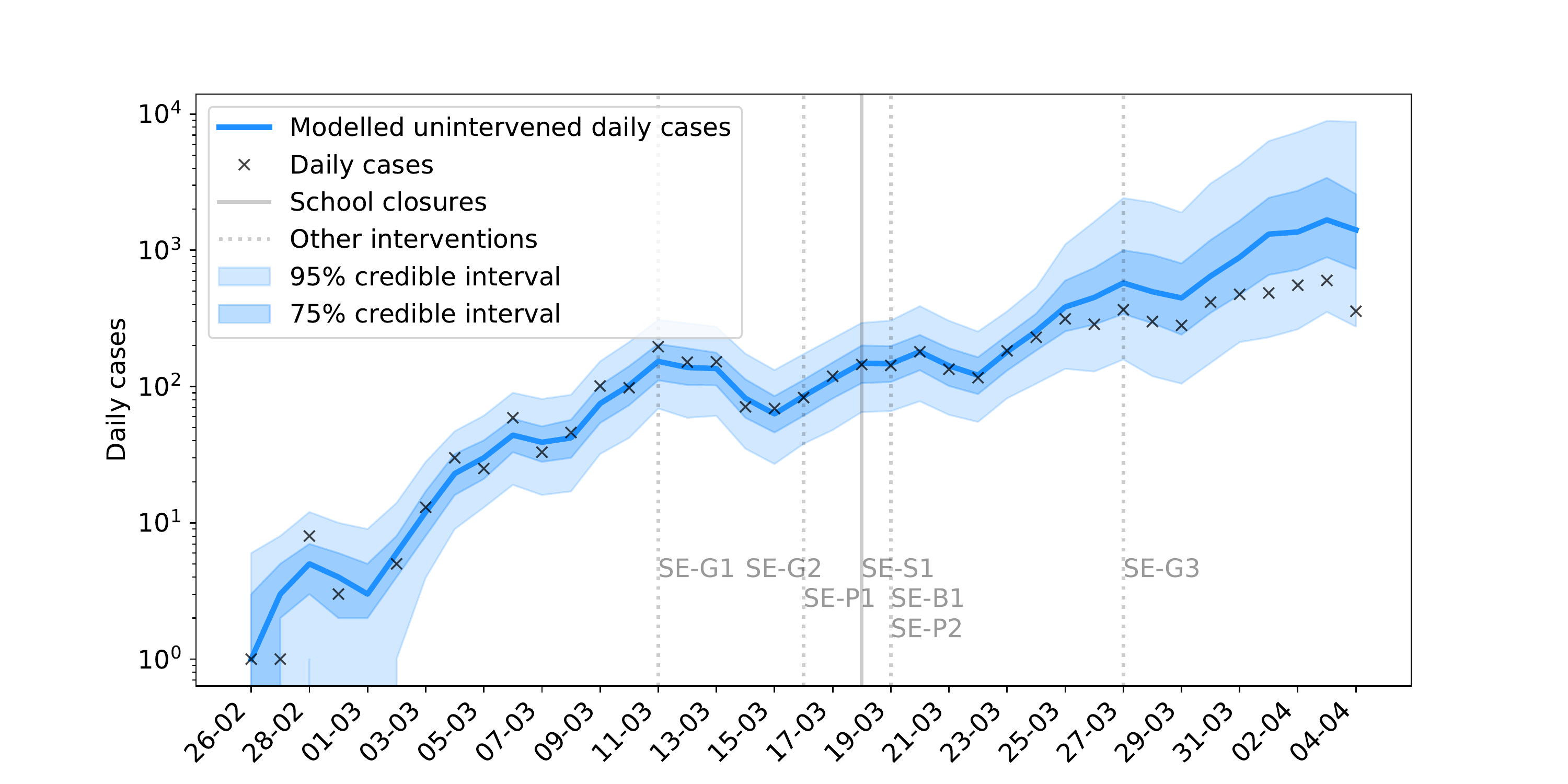}
\caption{\color{Gray}Modelled and observed daily cases in Sweden.} 
\label{SE_close}
\end{figure}

Sweden’s school closures were less restrictive than other countries', with only educational establishments for students aged 16 or over being required to close. Despite no official nationwide closing of primary or secondary schools in Sweden, there were local closures in response to outbreaks within the community. There is no evidence of a sustained reduction in the growth rate within a time period attributable to school closures (Figure \ref{SE_close}). It is notable, however, that the limited closures in Sweden were imposed in the absence of large-scale social restrictions. 
This indicates that school closures affecting older students without more widespread social interventions are unlikely to have significant national effects, and that school closures are most effective when implemented concurrently with other interventions.\newline
It is notable that there was an increase in weekly testing between March 30\textsuperscript{th} and April 6\textsuperscript{th}, which may have contributed to the apparent limited reduction in growth rate during this time. However, this falls outside of the time window in which we would expect to see a response attributable to school closures.

Sweden saw the following interventions introduced around the same time as school closures:
\begin{itemize}
    \item 11/03 - Banned gatherings of more than 500 people (SE-G1).
    \item 14/03 - Advice against non-essential travel (SE-G2).
    \item 16/03 - Social distancing advised but not enforced (SE-P1).
    \item 18/03 - \textbf{Closed all education for students aged 16 or over} (SE-S1).
    \item 19/03 - Restrictions on international travel (SE-B1); advice against national travel (SE-P2).
    \item 27/03 - Banned gatherings of more than 50 people (SE-G3). 
\end{itemize}

\subsection*{Reopening of schools}

\subsubsection*{Germany}
The following key interventions, sourced from \cite{WHONPI}, are possible confounders for the effects of school reopening:
\begin{itemize}
\item 20/04 - Opening of some retail venues (DE-R2).
\item 22/04 - Mandatory mask wearing in certain public spaces (DE-P3).
\item 27/04 - \textbf{Return of Year 10, final year exam students (ages 15, 18-19)} (DE-S3).
\item 29/04 - Extension of mask-wearing requirements (DE-P4).
\item 03/05 - Expiry of stay-at-home order (DE-G3).
\item 04/05 - \textbf{Return of Year 4 primary schools students (age 9)} (DE-S4); opening of retail (DE-R3) and public spaces (DE-P5).
\item 11/05 - \textbf{Return of primary and secondary school students (ages 9, 15, 17-19)} (DE-S5).
\item 15/05 - Relaxation of border controls (DE-B3).
\item 18/05 - \textbf{Staggered return of primary and secondary school students (ages 9-11, 15-19)} (DE-S6); meeting of two households allowed (DE-G4).
\item 29/05 - Gatherings of up to 10 people allowed (DE-G5).
\item 02/06 - Pubs reopen (DE-R4). 
\end{itemize}
Due to differing policies across German states, the dates of school reopening and the ages of students returning were variable. Where an age group appears over multiple dates, the return of students in this age group took place in different states. We present a summary of the overall national trend, as our data only distinguish between staff and students on the national scale. On three occasions the recorded cases were inconsistently reported, and values were imputed using cases reported on preceding and proceeding days. Our findings do not change significantly upon exclusion of these data points. We contrast these demographically specific findings by comparison with national hospital admissions.
\begin{figure}[H]
\centering
\includegraphics[width=130mm]{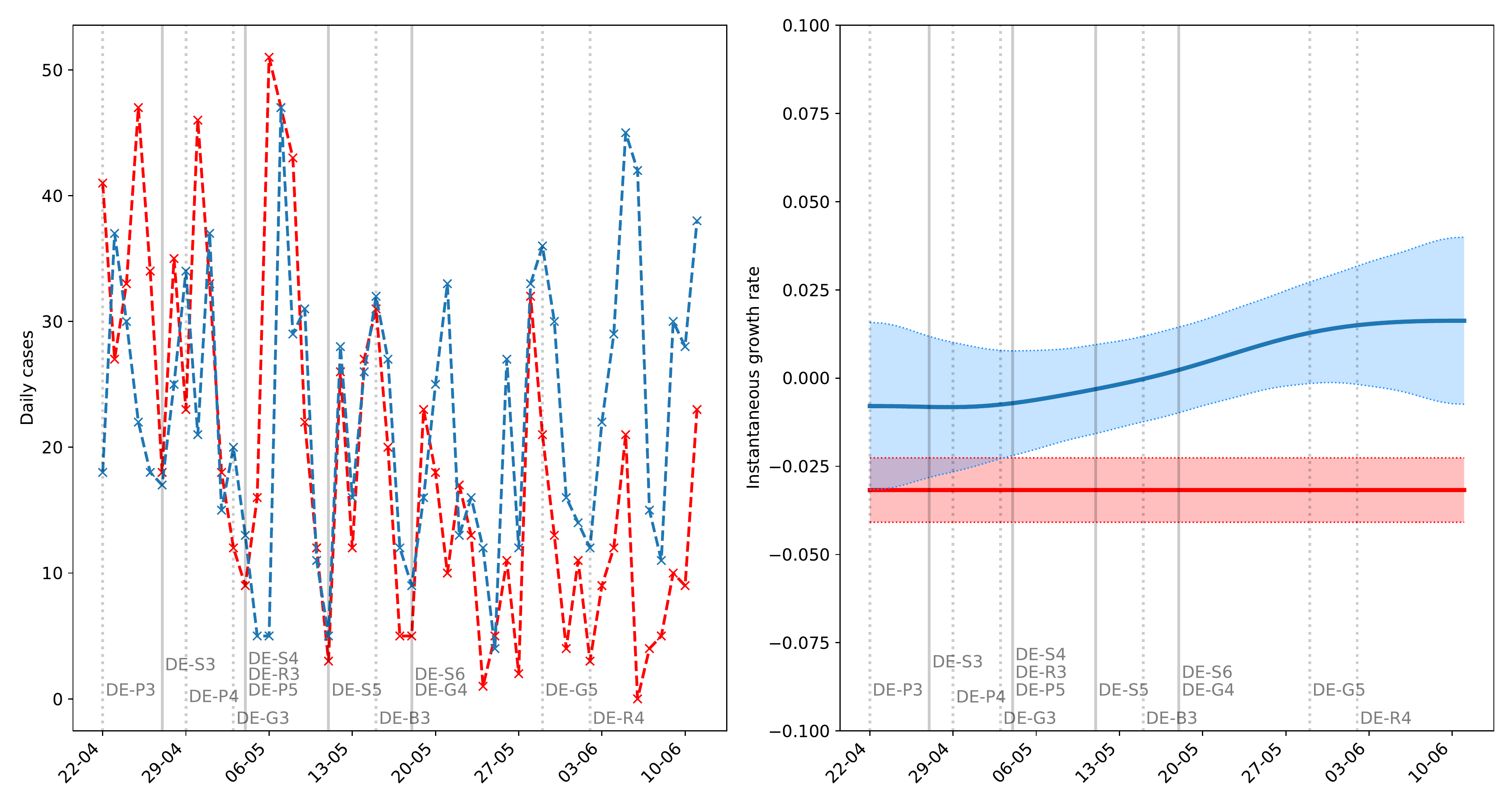}
\caption{\color{Gray}
Confirmed cases in staff (red) and students (blue) in schools, kindergartens, holiday camps, and other educational facilities for under-18s (age distribution not known) in Germany. Left shows daily new confirmed cases, and right shows the instantaneous growth rate (shaded regions are 95\% confidence intervals). Solid vertical lines indicate when students returned to school, and dashed lines indicate other changes to public measures. In April and early May with small numbers of primary school or exam students returning, there was no notable difference between the incidence among students and staff. Accounting for the lag time, the incidence among students was higher than that of staff following May 18\textsuperscript{th}.} 
\label{DE-schools}
\end{figure}

\begin{figure}[H]
\centering
\includegraphics[width=130mm]{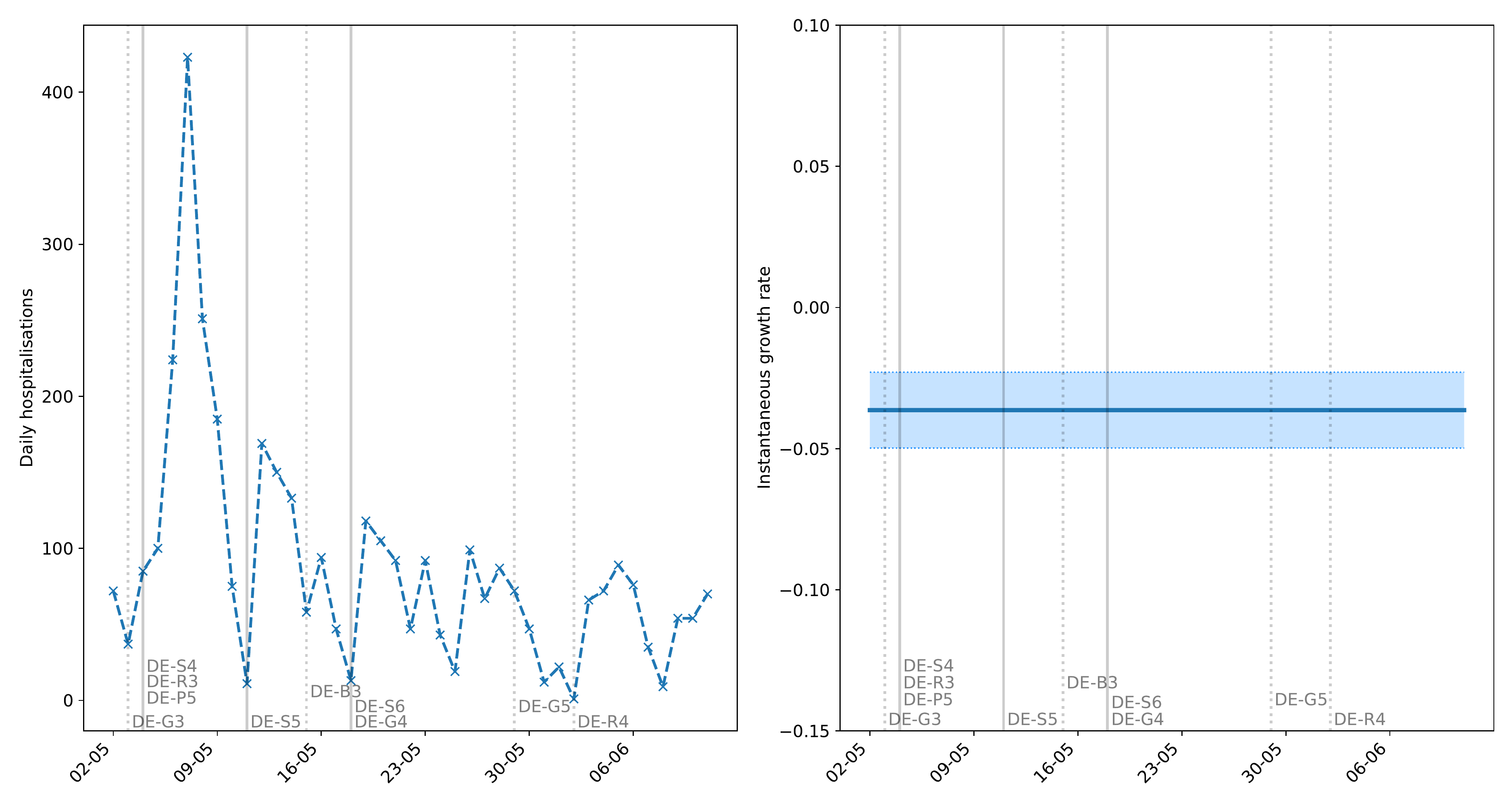}
\caption{\color{Gray}
Daily hospital admissions with COVID-19 in Germany, excluding those working in education, front-line healthcare workers, carers, catering, and hospitality, thus representing transmission in the general, average-exposure population. Left shows daily admissions, and right shows the instantaneous growth rate (shaded regions are 95\% confidence intervals). The continuing decline in admissions suggests that the return of younger (and exam) students did not present a statistically significant impact on the general hospitalised population.\newline
It is worth bearing in mind that hospital admissions lag further behind than confirmed cases. Additionally, since very few young people have been hospitalised, an additional generation time of 6 days \cite{Ferretti2020} may need to be added to this lag to account for students infecting older age groups.} 
\label{DE-hosp}
\end{figure}

The spike in daily cases observed around May 7-8\textsuperscript{th} (Figure \ref{DE-schools}, left) may be a result of increased presentation for testing following a national announcement of school reopening on May 4\textsuperscript{th} (allowing for testing delay), or increased community transmission following reopening of other parts of society which was subsequently contained.
Overall the incidence among staff decreased, which is supported by the growth rate among staff being negative. The incidence among students first decreased, and subsequently increased with a predominantly positive growth rate from the end of May (Figure \ref{DE-schools}, right).\newline\newline
The stable, low, values of the incidence and growth rate until the middle of May indicate that the return of final year and year 4 students either a) did not significantly increase transmission in schools or the community, or b) did increase transmission, but this was mitigated due to safety protocols of prevention and monitoring.
This observed effect is quite a strong signal as the daily case counts remain low even across a background of increased community transmission from late April onward with, for example, shops reopening. It is therefore reasonable to conclude that these age groups do not strongly increase transmission in a setting of effective social distancing.\newline\newline
However, the impact of most students returning to school from late May was different. In this time period, the incidence among staff qualitatively agreed with the national trend in hospitalisations (Figure \ref{DE-hosp}), i.e.\ staff did not immediately appear to be at greater risk following the return of more students. In contrast, the growth rate in student cases increased following May 18\textsuperscript{th}. The constant staff growth rate, and the small effect of the return of (mostly) younger years, suggests that the increased incidence may be due to a) increased transmission among older students, b) low feasibility of effective physical distancing in venues at full capacity, or c) a combination of these.

\subsubsection*{Denmark}
Schools reopened alongside the following key interventions sourced from \cite{WHONPI}:
\begin{itemize}
\item 08/04 - 7-day ban on gatherings of over 10 people (DK-G5).
\item 14/04 - Partial return of employees to work (DK-P2).
\item 15/04 - \textbf{Return of nursery, kindergarten, Year 0-5 primary school, and final year exam students (ages 0-12, 18-19)} (DK-S3).
\item 20/04 - Partial reopening of retail and small business (DK-R2).
\item 21/04 - Assemblies limited to 500 people (DK-G6).
\item 11/05 - Full reopening of shopping and retail (DK-R3).
\item 18/05 - \textbf{Return of Year 6-10 secondary school students (ages 12-16), and examinations requiring physical attendance} (DK-S4); restaurants and caf\'{e}s reopen (DK-R4); reopening of houses of worship (DK-P3).
\item 21/05 - Reopening of leisure and cultural facilities (DK-P4).
\item 25/05 - Relaxation of border restrictions with Nordic countries and Germany (DK-B2).
\item 27/05 - \textbf{Return of secondary school students (ages 16-18) and adult education} (DK-S5).
\end{itemize}

\begin{figure}[H]
\centering
\includegraphics[width=130mm]{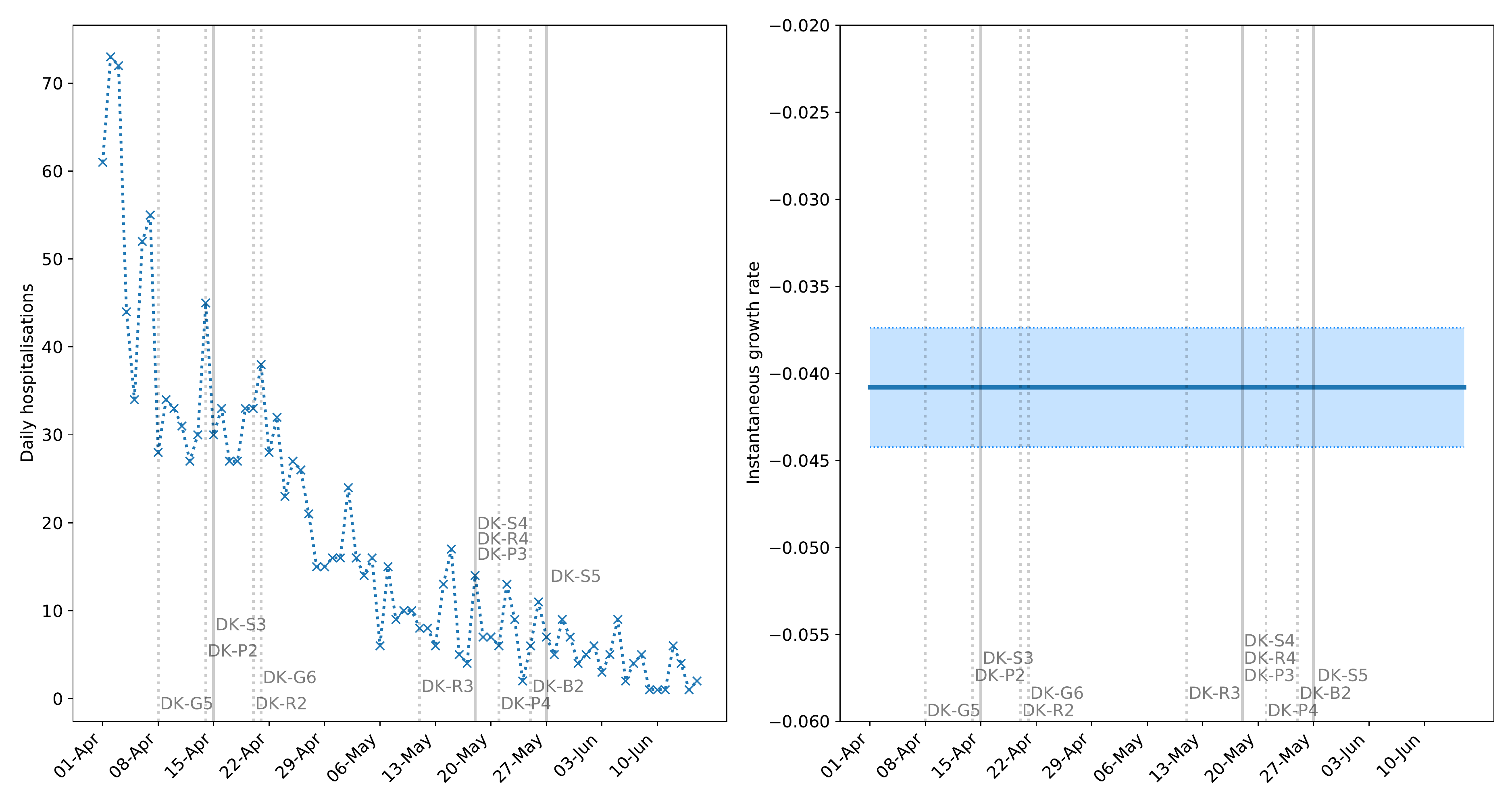}
\caption{\color{Gray}
Daily hospitalisations with COVID-19 in Denmark. Admissions are shown left, and right shows the instantaneous growth rate (shaded regions are 95\% confidence intervals). A longer lag time of 10-14 days is in effect from infection to hospitalisation \cite{StatensSerumInstitut2020}, with a further 6 days' generation time \cite{Ferretti2020} to account for subsequent infection generations due to the low hospitalisation rate among children.
Solid vertical lines indicate when students returned to school, and dashed lines indicate other interventions.
} 
\label{DK-hosp}
\end{figure}

There is no significant observable increase in the growth rate of hospital admissions following school reopening to younger years, even bearing in mind the subsequent reopening of some businesses (Figure \ref{DK-hosp}). The low growth rate and small relative number of admissions suggests that the return of younger years to school with social distancing did not contribute significantly to community transmission. The subsequent reopening stage on May 18\textsuperscript{th} also did not have a significant impact on hospital admissions, which we verify using confirmed cases (see Supplementary Material S5, School reopening analyses).\newline
These findings are further supported by a lower proportion of adults testing positive for COVID-19 among those working with children aged 0-16 than those working with students aged 16 or over (see Supplementary Material, S5) \cite{SSIsentinel}. However, these numbers alone do not distinguish between infection acquired from students and infection acquired elsewhere.

\subsubsection*{Norway}
The following events are possible confounders in the data, and key dates for school reopening sourced from \cite{WHONPI}:
\begin{itemize}
\item 01/04 - Exceptions made to entry restrictions (NO-B4).
\item 08/08 - Easing of entry restrictions from EEA workers (NO-B5).
\item 20/04 - \textbf{Return of kindergarten students (ages 1-5)} (NO-S2); travel to cabins allowed (NO-G5).
\item 27/04 - \textbf{Return of Year 1-4 (ages 6-10) and final year students (ages 18-19), vocational training, and higher education requiring physical attendance} (NO-S3); partial reopening of retail and small business (NO-R1).
\item 07/05 - Events, and some public sports and leisure facilities open, but limited to 50 people (NO-P1); group size for social gatherings increased from 5 to 20 people (NO-G6).
\item 11/05 - \textbf{Return of students aged 10-18 this week} (NO-S4); reopening of bingo halls and driving schools (NO-P2).
\item 12/05 - Easing on entry restrictions (NO-B6).
\end{itemize}

\begin{figure}[H]
\centering
\includegraphics[width=130mm]{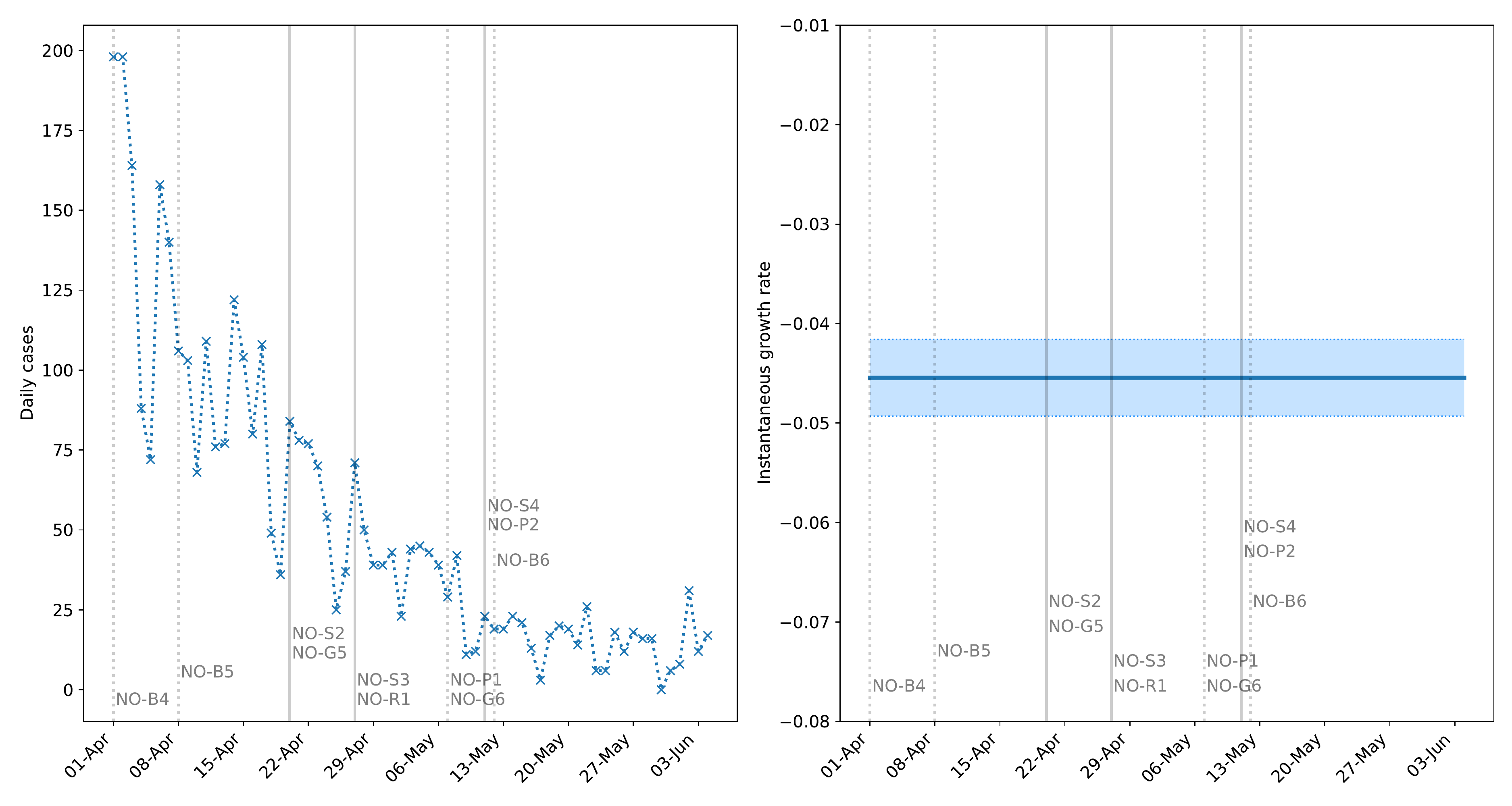}
\caption{\color{Gray}
Daily confirmed cases in Norway. The left panel shows new cases, and the right panel shows the instantaneous growth rate (shaded regions are 95\% confidence intervals). Solid vertical lines indicate when students returned to school, and dashed lines indicate other interventions. 
} 
\label{NO-cases}
\end{figure}

There is no notable change in the growth rate, even following the return of students in May (Figure \ref{NO-cases}). The consistently negative growth rate and small number of cases indicate that the return of most students to school (with social distancing) did not contribute significantly to community transmission. However, this effect is likely conditional on high levels of testing, with very low community transmission.

\section*{Discussion}

Decoupling the effect of school closure, and subsequent reopening, from other interventions is not straightforward. This work does not claim to have achieved this; however, the consistency of the signal across regions with different intervention timelines suggests a distinct effect of school closure on the subsequent growth in cases. The consistently lower post-intervention growth rates in German states when compared to the modelled scenarios with no interventions (see Table \ref{tab1}) suggests that school closures contributed to reducing the epidemic growth rate. School interventions were typically followed by a response in the data approximately 7 days later (corresponding to approximately 1.2 generations). Table \ref{tab1} shows that this lag time was comparable across states which closed schools at different times. High school students sitting their final examinations under social distancing does not appear to have significantly impacted case numbers. Sweden implemented partial school closures which affected students aged 16 or above. However, there is no evidence to suggest that this intervention affects the later daily incidence within the expected time frame. These findings are consistent with earlier works suggesting that school closure in isolation is insufficient to prevent the spread of COVID-19 \cite{Chen2020, Viner2020, Zhang2020a}. The evidence for the impact of school closures on growth rates in Norway and Denmark is more limited. While there was a reduction in growth rate of hospitalised cases after school closures, it has not been possible to link this effect with school closures.\newline

While school closures are often among the first implemented control measures, school reopenings are typically staggered with other eased restrictions, often with a small initial cohort of returning students. Since, to our knowledge, no stringent restrictions were introduced to compensate for the additional transmission risk due to school reopening, a lack of signal in the growth rate after reopening would be indicative that schools do not contribute substantially to community transmission. From our analysis, the reopening of schools to younger year groups and exam students in Germany, Denmark and Norway has not resulted in a significant increase in the growth rate. The return of all students (up to age 16 in Denmark) does not appear to have increased transmission in Denmark and Norway. However, the added return of most (primarily older) students in Germany has increased transmission among students, but not staff. It is unclear whether older students transmit more, or if physical distancing is practically unfeasible in classrooms at high capacity. The distinction between the impact of younger and older students is echoed in other modelling studies \cite{Domenico2020, Keeling2020}. Although our findings cannot be interpreted as causal links between individual interventions and changes in national case numbers, they represent a realistic assessment of the effects of school reopening in their natural context of wider societal interventions.\newline

Our findings are not without limitations. The presence (or lack) of signals in the data following school interventions are limited by the reliability of the available data. We have worked with  hospital admissions data, as they are less affected by variable testing, while bearing in mind that hospitalisations only affect a subset of the infected population. Where these data were unavailable, we have considered confirmed cases while monitoring the degree of testing in place to ensure such numbers were indicative. \newline

The GP regression method allows one to account for differences between the simulated epidemic trajectories from the ODE model and the observed cases. However, the fact that closures occurred very early on in the epidemic means that the GP method often had to be trained on a limited number of data points.\newline
Since the instantaneous growth rate relies on the derivative of splines, it is subject to increased error at the boundaries of the data. However, the observed signals are qualitatively robust to this limitation. Due to the noisiness of some data streams from relatively low incidence following mass quarantine, the values of the instantaneous growth rate should be taken as a quantification of the trend in incidence rather than the true value on any given day.\newline

The data have generally not made it possible to account for inevitable geographic variability, the age distribution of those studied, and their occupation (i.e.\ likelihood of exposure to infected individuals) in our analyses. The analysis of German school reopening, particularly the comparison of staff and student infections, warns against the reliability of using national-level data to understand the immediate effects or impact on a single population. Rather, such impact may only become visible in national data in subsequent generations. We must therefore be concerned not only with the lag time from intervention to a signal in the data, e.g.\ 7 days in Germany, but also with the following generation of infections.\newline

Our analysis is restricted to countries with high monitoring and intervention efficacy (including but not limited to high testing, tracing, and adherence to isolation), hence great care should be taken when translating our findings on the impact of school reopening to other nations. For instance, continued low incidence following the return of younger students does not imply that such a measure is inherently safe, but may rather be a result of the successful implementation of a complete system of monitoring and interventions jointly with low daily incidence, as observed in Denmark and Norway. In many instances, the students were spread over more classrooms, with greater levels of physical distancing from each other and teachers, conditions which are not always practically feasible.\newline
Caution is warranted surrounding the return of older students, in particular regarding the likely increased number of crowded classrooms, as well as their added impact to community transmission. The correlation with community transmission is particularly clear in Germany, with confirmed cases increasing among students, and the halted decay in hospital admissions on the national level. While all three countries seem to be effectively managing transmission, the volatility of new German hospital admissions warns that a failure in control, or a sudden spike in cases, will likely have a stronger effect in Germany than it would have in Denmark or Norway. Key to this observation is the aforementioned delay before which the ripple effects of school reopening will travel from students to the general population. Furthermore, we highlight that the tenuous balance (net zero growth in June) in Germany exists despite a swift and robust test and trace infrastructure and school-level stratified monitoring. We question the possibility of an equally effective reopening in countries with a monitoring process reliant on national-level incidence data. The swiftness and effectiveness of targeted interventions become increasingly crucial as the daily incidence increases, due to the correspondingly greater challenges presented in managing the localised outbreaks across e.g.\ reopened schools.\newline

Policy-makers should carefully consider their nations' respective capacities and associated effectiveness of infection management before considering a partial or full reopening of schools. Our findings suggest a small, strategically chosen, proportion of students to return in the first instance, with dedicated testing and monitoring of cases among staff and students (over time scales where the effect can be assessed). Any significant return of students to schools, particularly in countries with a high incidence, should not be considered unless an infrastructure is in place which would be able to swiftly identify and isolate most new cases as they appear \cite{Panovska-Griffiths2020}. Such a strategy may not be feasible if the community incidence is too high.\newline
When used in conjunction with household transmission models, and knowledge of the length of immunity associated with SARS-CoV-2, our findings may be used to inform age-targeted vaccine allocation protocols.

\subsection*{Authors' contributions}
JS modelled school closure from early data. HS analysed school reopening from late data. SG, LP, JS and HS drafted the manuscript. SG implemented the Gaussian process model. All authors contributed to the writing of the manuscript.

\subsection*{Funding}
HS and LP are funded by the Wellcome Trust and the Royal Society (grant 202562/Z/16/Z). FS is funded by the CIHR 2019 Novel Coronavirus (COVID-19) rapid research program. LP and FS are also supported by the UKRI through the JUNIPER modelling consortium (grant number MR/V038613/1). SG was supported by the Medical Research Council (Unit programme number MC UU 00002/11). JS and TF are funded through the Department for Health and Social Care grant in aid funding to Public Health England.

\subsection*{Acknowledgements}
We gratefully acknowledge Thomas House, Ian Hall and 
Pantelis Samartsidis for useful discussions and contribution to the methodology. In particular we acknowledge the work of Ian Hall in developing the GAM approach with a day-of-week effect.

\subsection*{Competing interests}
The authors declare no competing interests.

\subsection*{Data accessibility}
All data used in the production of this work, including the relevant scripts, can be found in the electronic supplementary material and online at \texttt{https://github.com/HelenaStage/COVIDschools}.

\subsection*{Ethical approval}
This study does not require ethical approval, as the only data used are publicly available.
\bibliography{library}

\bibliographystyle{unsrt}
\newpage
{\huge{\textbf{Supplementary Material}}}

\setcounter{figure}{0}
\renewcommand{\thefigure}{S\arabic{figure}}
\setcounter{table}{0}
\renewcommand{\thetable}{S\arabic{table}}
\setcounter{equation}{0}
\renewcommand{\theequation}{S\arabic{equation}}
\setcounter{section}{0}
\renewcommand{\thesection}{S\arabic{section}}

\section{Data availability}
The data streams used to carry out the analysis of this work for each country can be found in the following sources:
\begin{itemize}
    \item Denmark: daily tests, hospital admissions (used for school closure and reopening), and confirmed cases (used for school reopening) from the National Serum Institute \cite{SSIdata}. Sentinel survey among educational staff is from \cite{SSIsentinel}.
    \item Germany: daily (and weekly) tests, hospital admissions (used for school reopening), and confirmed cases (state-level used for school closure, and separated by staff and students used for school reopening) from the Robert Koch Institute \cite{RKIdata, rkidash}.
    \item Norway: daily tests, hospital admissions (used for school closure), and confirmed cases (used for school reopening) from the Norwegian Institute of Public Health \cite{NIPHdata}.
    \item Sweden: weekly tests, and confirmed cases (used for school closure) from the Public Health Agency of Sweden \cite{FHMdata}.
\end{itemize}

Note that some of these sources are updated over time, so that the numbers used in these analyses may not correspond exactly to those reported at the time of reading. Where we suspect this to be the case, our analysed data are enclosed with the relevant scripts.\newline
In the main manuscript and the analyses below, we highlight key interventions and measures concurrent with school closures and reopening. A comprehensive overview of these can be found in \cite{WHONPI} and sources therein.\newline

All data sources present their own biases, even beyond whether someone is suspected or confirmed as being infected with COVID-19. The available tools, procedures, and protocols for identifying and reporting cases have all changed over the course of this pandemic.\newline
Hospitalisations only capture those cases sick enough that they require medical attention, and is thus a biased representation of transmission towards the elderly and those with underlying health conditions. Furthermore, there are differences in reporting due to personal judgements from individual doctors and nurses. For example, the correct classification of nosocomial cases has been unclear in many cases. Similar challenges arise for patients admitted to hospital for a different treatment, but who also happened to be infected with SARS-CoV-2 with mild symptoms.\newline
Confirmed cases are strongly biased by testing; both the availability of the tests, their logistical deployment, and the laboratory capacity to analyse them. Further biases are present in the populations these are offered to: front-line healthcare staff, essential workers, care home residents, etc. This is further complicated if the allocation to various groups changes over time. Another bias in the data is what makes the general public eligible for a test: the number and severity of symptoms, their age, known recent contacts, and so on. If tests are only offered to symptomatic cases, the results will be biased away from younger people, as we know they are disproportionately likely to be asymptomatic or only mildly symptomatic. 

\section{Numerical methods}
\subsection{Overview of the method for counterfactual projection}

We adopt a two-step approach for generating the counterfactual projection. The first step consists in fitting a simple SEIR-type ODE-based compartmental model, using Approximate Bayesian Computation (ABC). Because we assume a constant transmission parameter, with the possible exception of a transient initial phase when the initial numbers propagate through the compartments, the model trajectories are characterised by exponential growth and cannot account for the impact of interventions prior to school closure (potentially deviating from exponential growth), nor for weekend effects. Therefore, in a second post-processing step we use a posterior sample of these trajectories as inputs to a Bayesian regression model that is in turn fitted to the same data. This post-processing step captures at least part of the discrepancy between the data and the smooth simulated ODE trajectories. 

For this regression model, we use a latent Gaussian process (GP) with a Negative Binomial likelihood. The choice of the Negative Binomial likelihood is dictated by the overdispersion inherent in the noisy incidence data, and hence captures the aleatoric uncertainty in the observational noise. The use of Gaussian processes allows us to specify priors on function spaces \cite{Rasmussen2006} for the Bayesian regression. The use of a Bayesian framework guarantees that not only a best fitting map linking the ODE trajectories and the data is used, but the entire posterior distribution of possible maps is explored. This properly quantifies and propagates both the uncertainty in the random observational noise, captured by the Negative Binomial likelihood, and the uncertainty in how well we are able to learn the structural discrepancy between the ODE trajectories and the data, which in this work is mainly due to weekend effects and possible systematic deviation from pure exponential growth before school closure can have any effect (most visible in the case of Rhineland-Palatinate, Figures \ref{main_closing} and \ref{GP_example}).

Once the GP is trained from the ODE trajectories and the data on the fitting interval (in the main results, up to 5 days after school closure, but sensitivity to this choice is explored in Table \ref{tab: sensitivity}), the same ODE trajectories past the end of the fitting interval are used by the GP to generate the counterfactual. In doing so, we assume that both the structural discrepancy between the ODE trajectories learnt by the GP and the overdispersion in the Negative Binomial observational noise remain unchanged within the counterfactual horizon.

\begin{figure}
	\centering
	\includegraphics[width=0.5\textwidth, height=0.45\textheight,keepaspectratio=true]{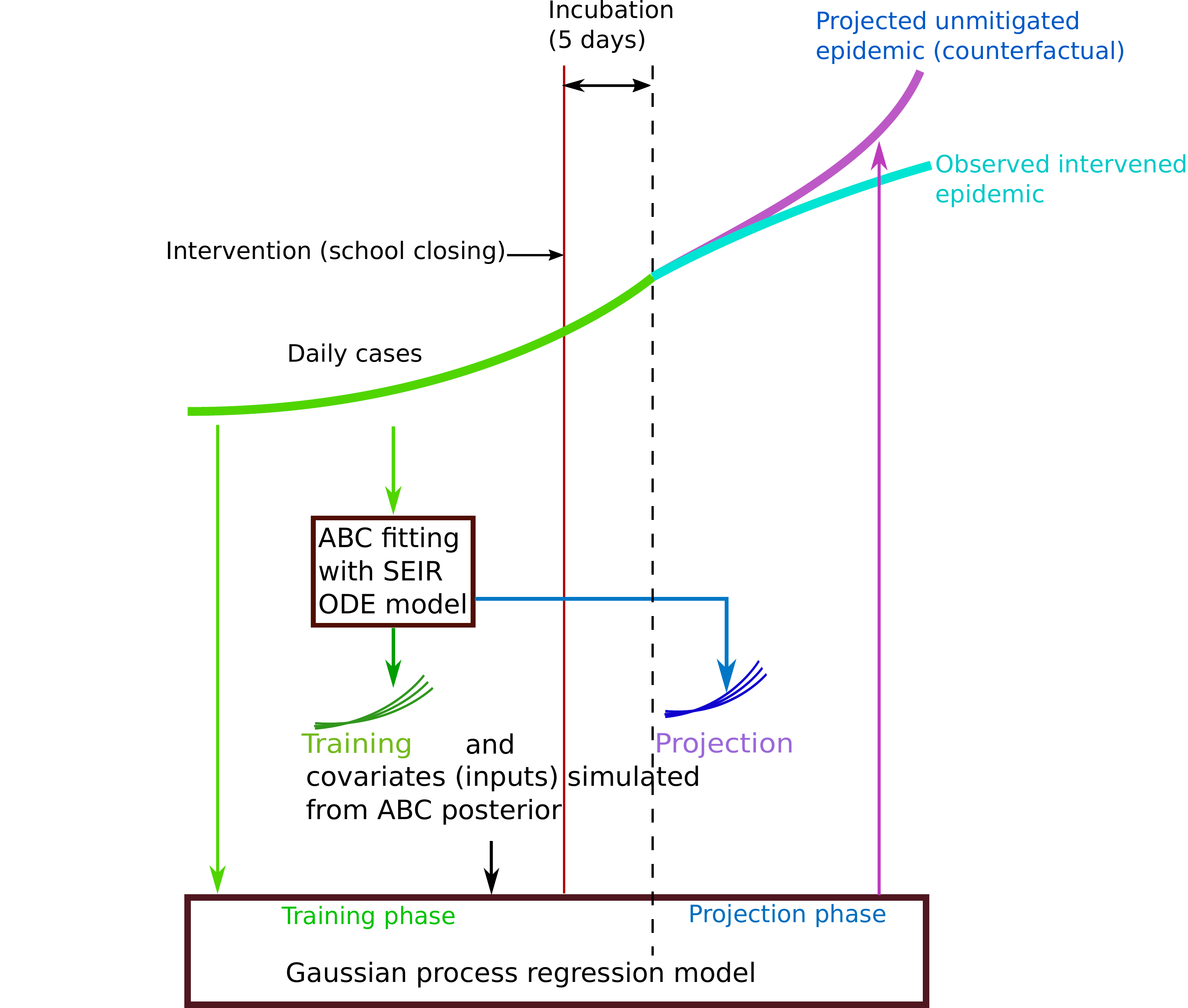}
	\caption{\label{fig:method-overview}\color{Gray}
		A cartoon illustration of the various steps that constitute our method for drawing counterfactual projections of an unmitigated epidemic.
	}
\end{figure}

Figure \ref{fig:method-overview} presents an illustration of the various steps that constitute our method. 
Further details of the method are described in detail in the following sections.

\subsection{The compartmental ODE model}

\begin{figure}[H]
\centering
\includegraphics[width=100mm]{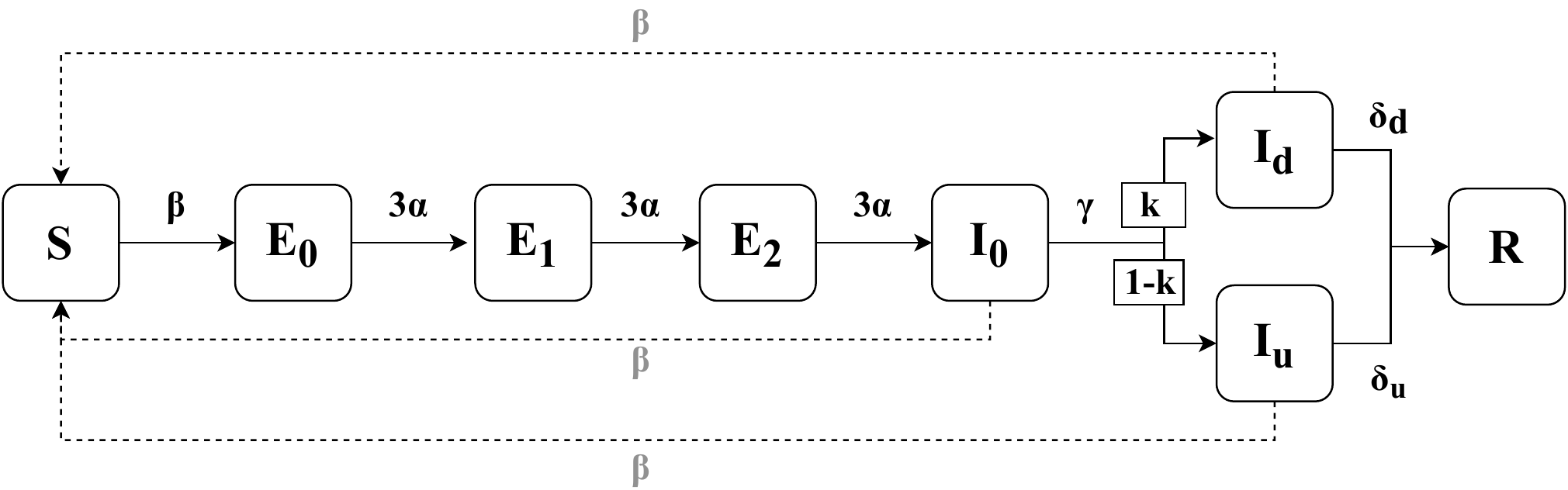}
\caption{\color{Gray}
The epidemic model used to simulate cases. The model uses multiple exposed compartments ($E_0, E_1, E_2$) to account for an Erlang-distributed incubation period.} 
\label{flowchart} 
\end{figure}

The compartmental model used to generate sample trajectories is outlined in Figure \ref{flowchart}. Multiple compartments have been used for the exposed ($E$) populations to model an Erlang-distributed incubation period compatible with available estimates of the mean and standard deviation of its duration \cite{Pellis2020a}. The early infection compartment $I_0$ splits into detected ($I_d$) and undetected ($I_u$) infectious populations. The same model is used for hospitalisations, with hospitalisations taking the place of $I_d$, and non-hospitalised cases taking $I_u$. A higher variability, possibly country-dependent, in the time from onset of symptoms to detection/hospitalisation \cite{Pellis2020a}, and limited knowledge on the duration of the infectious period and non-modelled pathways of hospitalised cases suggest a single compartment (i.e.\ exponential holding time) for these states is a reasonable and parsimonious choice. 

Denoting by $I = I_0+I_d+I_u$, the model equations read
\begin{align}
 \label{S}
    \frac{dS}{dt} &= -\beta S \frac{I}{N_0} \\
 \label{E0}
    \frac{dE_0}{dt} &= \beta S \frac{I}{N_0} - 3\alpha E_0 \\
 \label{E1}
    \frac{dE_1}{dt} &= 3\alpha (E_0 - E_1) \\
\label{E2}
    \frac{dE_2}{dt} &= 3\alpha (E_1 - E_2) \\
 \label{I0}
    \frac{dI_0}{dt} &= 3\alpha E_2 - \gamma I_0 \\
 \label{Id}
    \frac{dI_d}{dt} &= k\gamma  I_0 - \delta_d I_d \\
 \label{Iu}
    \frac{dI_u}{dt} &= (1-k)\gamma I_0 - \delta_u I_u \\
 \label{R}
    \frac{dR}{dt} &= \delta_d I_d + \delta_u I_u
\end{align}

\begin{table}
\centering
\begin{tabular}{@{}cccc@{}}
\toprule
\textbf{Parameter} & \textbf{Interpretation} & \textbf{Value} & \textbf{Priors}\\ 
\cmidrule{1-4}
$N_0$ & Population & Various & N/A\\
$\beta$ & Transmission rate & Inferred in ABC fitting & $0.25\leq\beta\leq4$\\
$\alpha$ & Rate of transition through latent period & $1/4.8 \text{ days}^{-1}$ \cite{Pellis2020a} & N/A\\
$k$ & Proportion of infectious individuals who are detected & Inferred in ABC fitting & $0 \leq k \leq 1$\\
$\gamma$ & Rate of transition through infectious period & Inferred in ABC fitting & $0.05 \leq \gamma \leq 1$\\
$\delta_d$ & Rate of removal of detected infectious individuals & Inferred in ABC fitting & $0.03 \leq \delta_d \leq 1$ \\\ 
$\delta_u$ & Rate of removal of undetected infectious individuals & Inferred in ABC fitting & $0.03 \leq \delta_u \leq 1$\\ 
\bottomrule
\end{tabular}
\caption{\label{tab: priors}\color{Gray}
A summary of the parameters used in the ODE model. The only parameters not inferred are the rate of transition through the latent period, $\alpha$, and the population, $N_0$. Population data are taken from various official sources \cite{DEpop, DKpop, NOpop, SEpop}. The priors for the respective parameters are chosen as uniform distribution with ranges furnished in the last column.} 
\end{table}
The data for Germany are only available starting from the 4\textsuperscript{th} of March, by which time, most states had already seen a number of cases. As such, we have inferred the initial values for states $\Phi = \{E_0(0), E_1(0), E_2(0), I_0(0), I_d(0),I_u(0)\}$. For consistency the same initial conditions have been fitted for Sweden, Denmark and Norway. 

Denoting the daily symptomatic case count as $D(t)$, we set the priors on the initial states $\Phi_i \in \Phi$, $i = 1,..,6$, as $D(0)/4 \leq \Phi_i \leq 4 \cdot D(0)$ for all $i$. 

We denote the unknown parameter vector of the model as: $\thb = (\beta, \gamma, k, \delta_d, \delta_u; \Phi)$. The parameter $\alpha$ is not estimated in this method, and is instead taken from the mean incubation period \cite{Pellis2020a} to be $\alpha=1/4.8$ days$^{-1}$. The interpretations of these parameters are provided in Table \ref{tab: priors}.

\subsection{Step 1: Approximate Bayesian Computation for fitting the ODE}
We used the Approximate Bayesian Computation (ABC) method to carry out Bayesian inference for estimating the ODE parameters, which we describe next. We denote by $y(t)$ the daily cases observed on day $t$ and we regard them as the noisy observation corresponding to the unobserved true number of symptomatic cases $D(t)$ on the same day. Having put a prior distribution $\pi(\thb)$ (see Table \ref{tab: priors} for a description of the priors) on the parameter vector we can obtain the posterior distribution, using the Bayes' theorem as:
\begin{equation}
    p(\thb|\bv{y}) \propto \pi(\thb) p(\bv{y}|\bv{D}^{\thb},\thb),
\end{equation}
where $p(\bv{y}|\bv{D}^{\thb},\thb)$ is the likelihood term. The vectors $\bv{y}=[y(1), \ldots, y(T)]$ and $\bv{D}^{\thb}=[D(1), \ldots, D(T)]$ denote the observed data and corresponding simulation from the model for a period of $t=1, \ldots, T$ days, given the parameters $\thb$. 

In ABC the likelihood is replaced by a distance function $d(\bv{y},\bv{D}^{\thb})$, between the data, $\bv{y}$, and the model simulation, $\bv{D}^{\thb}$. The ABC posterior given by
\begin{equation}
   p_{\epsilon}(\thb|\bv{y}):=p(\thb|\mathbbm{1}\{ d(\bv{y},\bv{D}^{\thb}) \leq \epsilon \}),
\end{equation}
which depends upon a user defined tolerance threshold $\epsilon$. We can obtain the exact posterior when $\epsilon \rightarrow 0$. $\mathbbm{1}(\cdot)$ denotes the indicator function. Rather than using a single tolerance threshold, we used the ABC-SMC \cite{Toni} algorithm that generates a sequence of intermediate distributions, each corresponding to a decreasing tolerance threshold. The final distribution in the sequence is the desired ABC posterior.

We employ a Negative Binomial probability mass function to calculate the probability of observing the data when the unobserved true values are given by the simulated trajectories. That is, for each day $t$ in a set of $T$ data points used for the fit, given the observed value $y(t) \in \bv{y}$ and the modelled value $\hat{y}(t) \in \bv{D}^{\thb}$,
the probability mass function is given by:
\begin{equation}
    P(y(t);\hat{y}(t),k) = \frac{\Gamma(k+y(t))}{\Gamma(k)y(t)!}\left(\frac{k}{k+\hat{y}(t)}\right)^k\left(\frac{\hat{y}(t)}{k+\hat{y}(t)}\right)^{y(t)},
\end{equation}
where $k$ is the over-dispersion parameter, which is also estimated from the data set. The distance function is then defined as:
\begin{equation}
d(\bv{y},\bv{C}^{\thb}) = -\sum_{t=1}^T\ln(P(y(t);\hat{y}(t),k)).
\end{equation}

\subsubsection{Generating covariates for the GP regression}
After obtaining the posterior distribution we generate simulated epidemic curves $\ipre:=\left[D^{\thb^*}(1), \ldots, D^{\thb^*}(T)\right]$, by solving the compartmental model using samples from the ABC posterior $\thb^* \sim p_{\epsilon}(\thb|\bv{y})$ for the period up to 5 days after school closures. Similarly, we generate posterior predictive trajectories $\ipost$ for the post-intervention period of $T^*$ days. This completes the first step of our method. We then select $M=15$ simulated epidemic curves  $(\ipre,\ipost)$ evenly distributed over the Bayesian credible interval of the posterior predictive distribution of $(\ipre,\ipost)$, to be used as the respective training and projection covariates for the GP regression which constitutes the second step of our method. Essentially, with these $M$ covariates we are simply summarising the posterior predictive distribution at each time point. Note that these $M$ summaries are not treated as random variables but as fixed covariates. Testing of the method with different numbers of covariates ($M=5$, $M=40$ and $M=150$) has lead to qualitatively and quantitatively similar results, suggesting the method is robust to the choice of $M$.

\subsection{Step 2: Learning discrepancy using GP regression}
In order to generate the unmitigated epidemic, the counterfactual, we first learn a map $\ipre \mapsto \bv{y}$, that represents the discrepancy as a nonlinear function. To achieve this we model the case data $\bv{y}$ prior to the intervention using a hierarchical latent variable formulation of GP regression with discrete outcomes. At the first level of this hierarchy we model each data-point $y(t)$, independently, using a Negative Binomial distribution given by
\begin{equation}
    y(t) \sim \operatorname{NegBin}(\mu(t), \eta),
\end{equation}
where $\eta$ is a dispersion parameter such that $\mathbb{E}[y(t)]=\mu(t)$ and $\operatorname{Var}[\bv{y}(t)]=\mu(t)(1+ \mu(t)/\eta)$. The mean $\bv{\mu}=[\mu(1),\ldots, \mu(T)]$ is in turn driven by a latent Gaussian process as:
\begin{equation}\label{eq: GP model}
    \log \bv{\mu} \sim \mathcal{GP}\big(\ipre \bv{\beta}+ \bv{b}, \cov(\ipre, \ipre^{'};\phb)\big),
\end{equation}
where we have used a linear regression form to describe the mean function of the GP with coefficients $\bv{\beta}$, an $M$-dimensional vector, and intercept term $\bv{b}$, a $T$-dimensional vector with elements set to $b$. This ensures that the GP, while extrapolating further away from the training data, follows the ODE predictions. $ \cov(\ipre, \ipre^{'};\phb)$ is the covariance function (also known as covariance \textit{kernel}) parameterised by $\phb$, the details of which are described in the next section. 

\subsubsection{Capturing the weekend effect}
Gaussian processes are considered to be a prior on function spaces \cite{Rasmussen2006}. The prior knowledge about a function modelled by the GP is generally introduced through the covariance function. For our problem we expect the function $\log \bv{\mu}$ to be a smooth yet periodic function. The periodicity assumption is introduced to cater for the weekend effect. To design such a prior we first introduce a new categorical \textit{day-of-week} covariate $\Wpre=(w(1), \ldots, w(T)), \quad w(t) \in \{ 1, \ldots, 7\}$, that indicates the day of a week for the $T$ consecutive training days, and $\Wpost$ for the forecasting days. We then replace the covariance of the GP, in Equation \eqref{eq: GP model}, with the sum of two separate covariances:
\begin{equation}
  \cov(\ipre, \ipre^{'}, \Wpre, \Wpre^{'};\phb_1,\phb_2) = \cov_1(\ipre, \ipre^{'};\phb_1) + \cov_2(\Wpre, \Wpre^{'};\phb_2),
\end{equation}
based on these different covariates, and having individual parameters $\phb_1, \phb_2$. By choosing valid covariance kernels components $\cov_1$ and $\cov_2$, it is ensured that we produce a valid overall covariance kernel $\cov$.

For the first component we chose the \textit{Matern} kernel \cite{Rasmussen2006} given by
\begin{equation}\label{eq:matern kernel}
    \cov_1(\ipre,\ipre^{'};\phb_1)= \alpha_1^2 \left(1 + \sum_{j=1}^M\frac{\sqrt{3(C_{pre_{j}} - C_{pre_{j}}^{'})^2}}{\rho_j}\right)
           \left(\sum_{j=1}^M \mathrm{exp}\left[ - \frac{\sqrt{3(C_{pre_{j}} - C_{pre_{j}}^{'})^2}}{\rho_j} \right]\right),
\end{equation}
where $M$ is the number of covariates, that is the number of simulated trajectories $\ipre$ generated using the ABC posterior as described previously. The parameter $\rho_j$ quantifies the characteristic length scale along the $j$\textsuperscript{th} covariate and $\alpha_1$ denotes the marginal variance of the GP prior. Together they constitute the parameter vector $\phb_1=\left[\alpha_1, \rho_1, \ldots, \rho_M\right]$ (note that $\rho_j$ is distinct from the quantity $\rho(t)$ used to estimate the instantaneous growth rate for school reopening).

For the second component, we chose the \textit{exponentiated quadratic} kernel:
\begin{equation}
    \cov_2(\Wpre, \Wpre^{'};\phb_2)=\alpha^2_2\exp\left(- \frac{\|\Wpre-\Wpre^{'}\|^2}{2\rho^2} \right),
\end{equation}
where $\phb_2 =(\alpha_2, \rho)$ are the covariance parameters.

\subsubsection{Estimation of the GP parameters using MCMC}
The latent GP represents the second level in the hierarchy and finally we place priors $p(\bv{\beta},b,\phb_1,\phb_2)$ on all the parameters to completely specify the generative model. Thus, the likelihood corresponding to all $T$ observations is given by
\begin{equation}
    p(\bv{y}|\ipre,\Wpre,\bv{\beta},b,\phb_1,\phb_2,\eta) = \Big(\prod_{t=1}^T \operatorname{NegBin}(e^{\mu(t)},\eta)\Big) \mathcal{N}(\bv{\mu};\ipre\bv{\beta} + \bv{b},\Cov_{TT}),
\end{equation}
where $\Cov_{TT}$ represents the covariance function (Eq.~\eqref{eq:matern kernel}) evaluated on all $T \times T$ pairs of the covariates $\ipre$ and $\Wpre$:
\begin{equation}
    \Cov_{TT} = \begin{pmatrix}
        \cov(\icone,\icone,\Wone,\Wone;\phb_1,\phb_2) &  \ldots & \cov(\icone,\icN, \Wone, \WN;\phb_1,\phb_2) \\
        \vdots   & \ddots & \vdots  \\
        \cov(\icN,\icone,\WN,\Wone;\phb_1,\phb_2) & \cdots & \cov(\icN,\icN,\WN,\WN;\phb_1,\phb_2) 
    \end{pmatrix}.
\end{equation}
We then combine the likelihood with the priors to obtain the posterior distribution of the parameters given by
\begin{equation}
    p(\bv{\beta},b,\phb_1,\phb_2,\eta|\bv{y})\propto p(\bv{y}|\ipre,\Wpre\bv{\beta},b,\phb_1,\phb_2)p(\bv{\beta},b,\phb_1,\phb_2,\eta).
\end{equation}
We used MCMC, applying the No-U-Turn (NUTS) \cite{hoffman2014no} algorithm provided by the PyMC3 package, to sample from this posterior. Note that to apply MCMC we need to sample the latent Gaussian process for the mean $\bv{\mu}$. It is well known (see \cite{Filippone}) that sampling both the GP parameters $\phb_1,\phb_2$ as well as the latent GP itself drastically reduces the efficiency of the MCMC algorithm and poor mixing is observed. For this reason a non-centred reparameterisation is used which gives the GP sample as:
\begin{equation}
    \bv{\mu} = \ipre\bv{\beta} + \bv{b} + \bv{L}_{TT}\bv{\epsilon},
\end{equation}
where, $\bv{L}_{TT}$ is obtained through the Cholesky decomposition of $\Cov_{TT}$ and $\bv{\epsilon}\in \mathbb{R}^T$ is a sample from the standard Normal density $\mathcal{N}(\bv{0},\bv{I})$ ($\bv{I}$ being the identity matrix).

\subsubsection{Projecting the counterfactual}
Having estimated the posterior $p(\bv{\beta},b,\phb_1,\phb_2|\bv{y})$ we can obtain the posterior predictive distribution of the daily cases, in the case of no intervention, $\bv{y^*}= [y^*(1), \ldots, y^*(T^*)]$, where $T^*$ is the number of days we want to project after the intervention date. This distribution is given by
\begin{equation}\label{eq:ppc}
    p(\bv{y^*}|\ipre,\ipost,\Wpre,\Wpost,\bv{y}) = \int \Big(\prod_{t=1}^{T^*} \operatorname{NegBin}(e^{\mu(t)},\eta)\Big)\mathcal{N}(\bv{\mu};\bv{m},\bv{V})p(\bv{\beta},b,\phb_1,\phb_2|\bv{y}) d\bv{\mu} d\bv{\beta} db d\phb_1d\phb_2d\eta,
\end{equation}
where the posterior mean $\bv{m}$ and variance $\bv{V}$ of the GP is given by \cite{Rasmussen2006}:
\begin{equation}\label{eq:gp mean var}
    \begin{aligned}
        \bv{m} &= (\ipost\bv{\beta} + \bv{b}) + \Cov_{T^{*}T}\Cov_{TT}^{-1}(\bv{y}-\ipre\bv{\beta} - \bv{b})\\
        \bv{V} &= \Cov_{T^{*}T^{*}}- \Cov_{T^{*}T}\Cov_{TT}^{-1}\Cov_{TT^{*}}.
    \end{aligned}
\end{equation}
$\Cov_{T^{*}T}$ and $\Cov_{TT^{*}}$ denote the $T^{*} \times T$ and $T \times T^{*}$ matrices of covariance function evaluations between the training (pre-intervention) $\ipre,\Wpre$ and projection (post-intervention) $\ipost,\Wpost$ covariates:
\begin{equation}\label{eq:cross-covariance-1}
    \Cov_{T^{*}T} =  \begin{pmatrix}
        \cov(\ivone,\icone,\Wvone,\Wone;\phb_1,\phb_2) &  \ldots & \cov(\ivone,\icN,\Wvone,\WN;\phb_1,\phb_2) \\
        \vdots   & \ddots & \vdots  \\
        \cov(\ivM,\icone,\WvM,\Wone;\phb_1,\phb_2) & \cdots & \cov(\ivM,\icN,\WvM,\WN;\phb_1,\phb_2) 
    \end{pmatrix},
\end{equation}
\begin{equation}\label{eq:cross-covariance-2}
    \Cov_{TT^{*}} =  \begin{pmatrix}
        \cov(\icone,\ivone,\Wone,\Wvone;\phb_1,\phb_2) &  \ldots & \cov(\icone,\ivM,\Wone,\WvM;\phb_1,\phb_2) \\
        \vdots   & \ddots & \vdots  \\
        \cov(\icN,\ivone,\WN,\Wvone;\phb_1,\phb_2) & \cdots & \cov(\icN,\ivM,\WN,\WvM;\phb_1,\phb_2) 
    \end{pmatrix},
\end{equation}
and $\Cov_{T^{*}T^{*}}$ is the covariance evaluated at the projection inputs $\ipost$ only:
\begin{equation}
   \Cov_{T^{*}T^{*}} =  \begin{pmatrix}
        \cov(\ivone,\ivone,\Wvone,\Wvone;\phb_1,\phb_2) &  \ldots & \cov(\ivone,\ivM,\Wvone,\WvM;\phb_1,\phb_2) \\
        \vdots   & \ddots & \vdots  \\
        \cov(\ivM,\ivone,\WvM,\Wvone;\phb_1,\phb_2) & \cdots & \cov(\ivM,\ivM,\WvM,\WvM;\phb_1,\phb_2) 
    \end{pmatrix}.
\end{equation}
We use Monte Carlo integration, using the samples of $\bv{\beta},b,\phb_1,\phb_2,\eta$ obtained through MCMC and samples of $\bv{\mu}$ generated using the aforementioned reparameterisation, to implicitly evaluate the desired posterior predictive distribution, given by Eq.~\eqref{eq:ppc}, and subsequently the counterfactual epidemic projections.

\subsection{Priors for the GP and MCMC details}
For each of the coefficients $\bv{\beta}$ and the intercept $b$ parameters we chose a Student's t-distribution with $5$ degrees of freedom. The mean is set to $0$ and the standard deviation is set to $1$. This was done following the recommendations in \cite{gelman2008weakly} for choosing priors for regression (since we modelled the GP's mean function using a linear regression form).

We used a half-Cauchy prior with scale set to $5$ for the marginal variances $\alpha_1,\alpha_2$, following recommendations in \cite{gelman2006prior}. For each characteristic length scale parameter we have placed a $\operatorname{Gamma}(2,1)$ prior.\newline
For the dispersion parameter, $\eta$, of the negative binomial we placed a $\operatorname{Gamma}(2,0.5)$ prior.

We have also used a logarithmic transform of $\ipre,\ipost$ before passing them to the GP. 

We ran NUTS with 2 chains for a total $1000$ iterations, where we discarded the first $500$ iterations as burnin (warmup). We set a target acceptance rate of $0.98$ for NUTS during the warmup phase iterations. As with the SEIR model fit, the GP model is trained with data up to 5 days after school closure.

\subsection{Projection comparison: the benefit of using the GP}
\begin{figure}[h!]
    \centering
    \begin{subfigure}[t]{0.5\textwidth}
        \centering
        \includegraphics[height=1.5in]{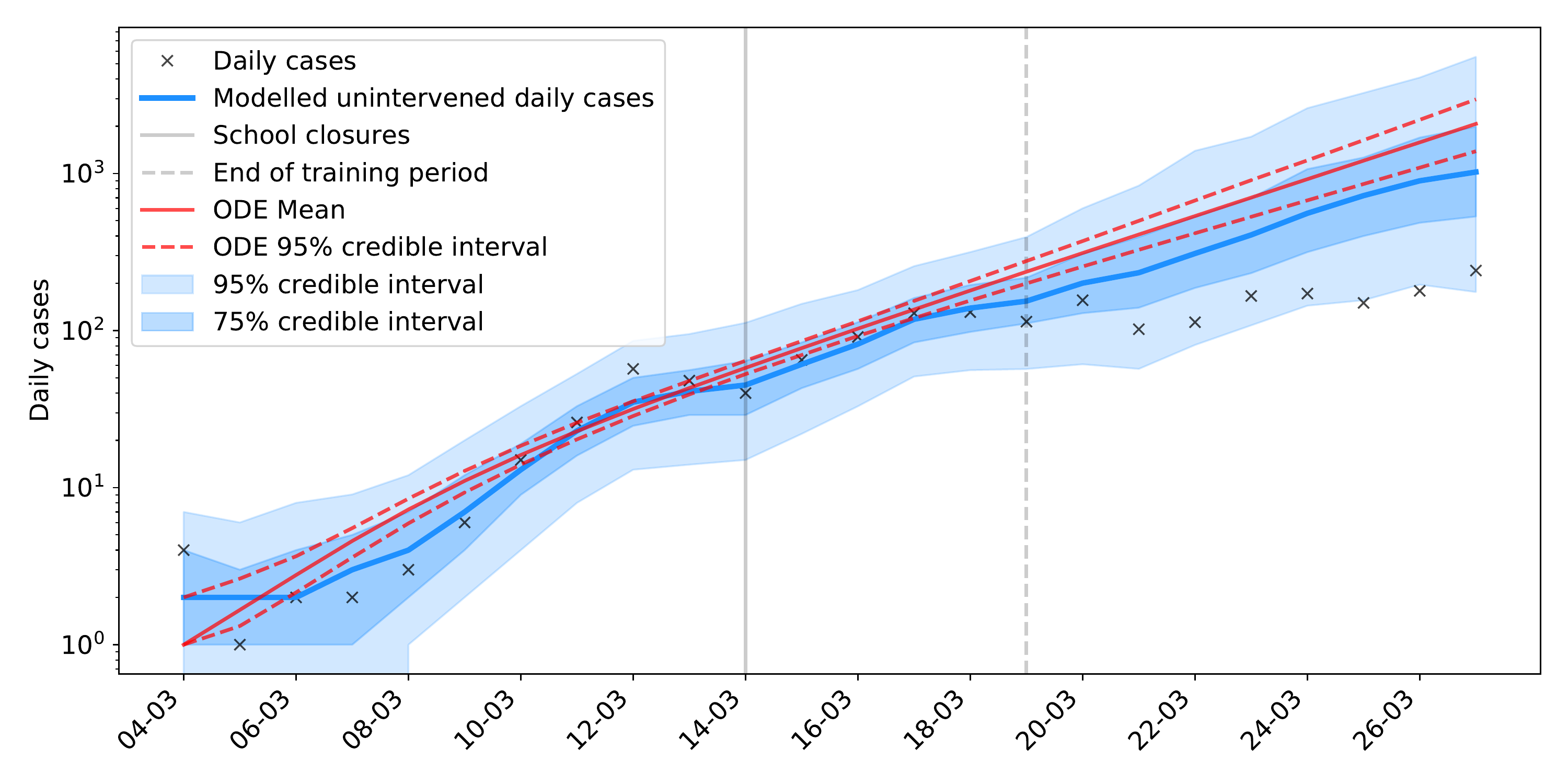}
        \caption{Gaussian Process and ABC fit for Rhineland-Palatinate.}
    \end{subfigure}%
    ~ 
    \begin{subfigure}[t]{0.5\textwidth}
        \centering
        \includegraphics[height=1.5in]{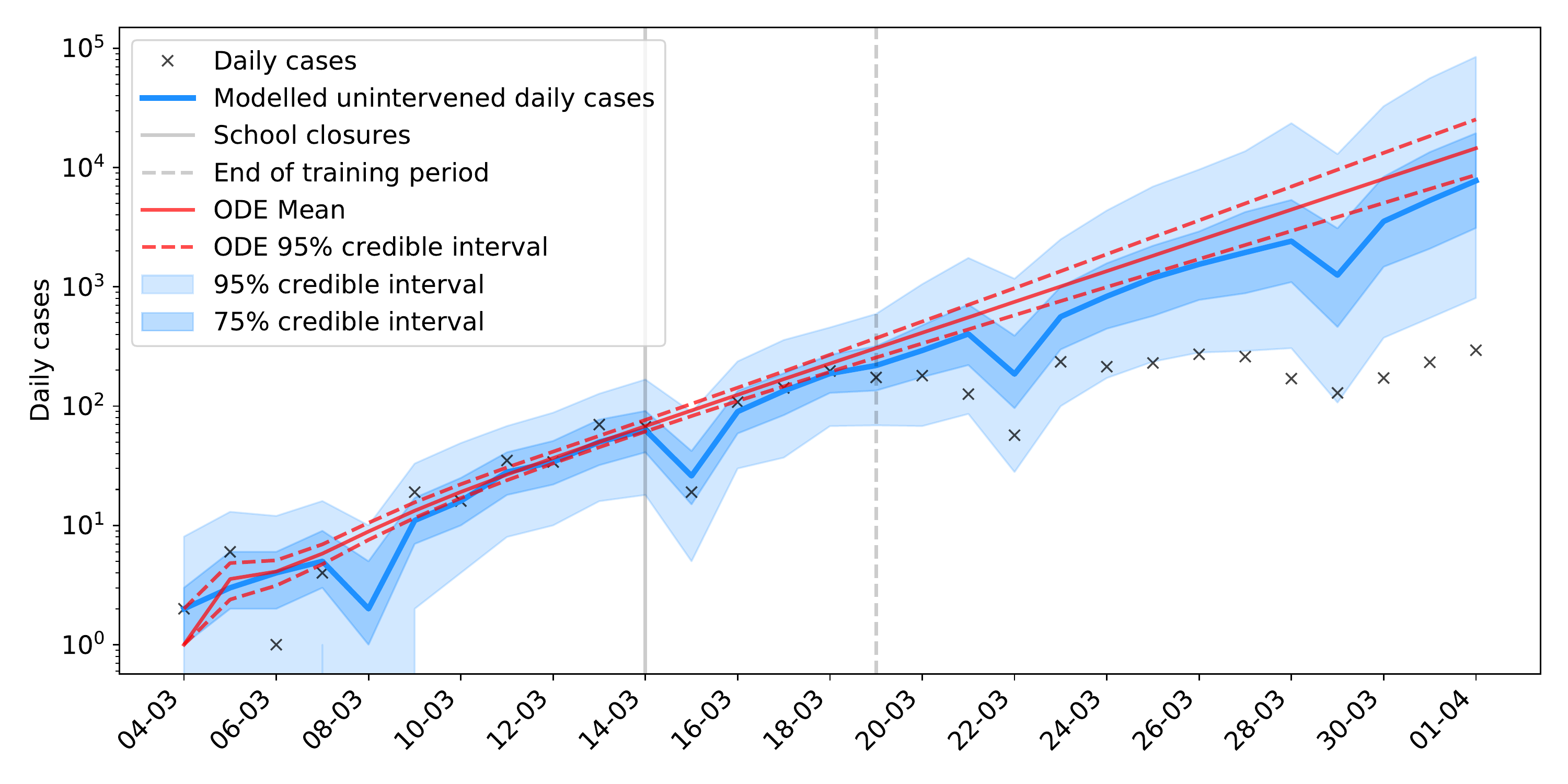}
        \caption{Gaussian process and ABC fit for Hesse.}
    \end{subfigure}
    \caption{\color{Gray}In (a) the ABC fitting process failed to account for a change in growth rate in the observed data directly before intervention. The GP method, however, is able to capture this change in the training period and adapt the projection accordingly. Similarly, in (b) the GP model is able to pick up both the slight change in growth rate and the strong weekend effect in the data, factors which are not addressed by the ABC fitting process.}
    \label{GP_example}
\end{figure}

Figure \ref{GP_example} shows the sample regression outputs in two German states, plotted alongside the mean and 95\% credible intervals of the ABC posterior projections (simulations from the fitted ODE model) used as covariates in the GP model. In both cases, the ODE model fails to capture the change in growth rate, as well as the weekend effect (which is visibly more pronounced in one state than the other). The post-processing with the GP regression compensates for this failure by learning the discrepancy as a function mapping the ODE simulation to the observations. 

\subsection{Sensitivity of the lag time to the training period}
Due to the noisiness of the data, our findings will inevitably be sensitive to the number of data points which we include in any analysis we do. This is most apparent in the fitting and training periods of the ABC and GP, which in this work extend to 5 days past school closure. This duration is chosen as it is the mean incubation period for SARS-CoV-2 \cite{Pellis2020a}, and thus we would not expect any signal during those days to be strongly attributable to school closures. Five days is thus our \textit{minimal lag time}, which acts as an input to our models. While a full sensitivity analysis of our modelling choices is beyond the scope of our investigation, we include some choice illustrations of the effects of changing this minimal lag time in Table \ref{tab: sensitivity}:

\begin{table}[H]
\ra{1.3}
\centering
\begin{tabular}{@{}lcccc@{}}
\toprule
\textbf{State} & \makecell{Minimal lag \\ time of 4 days} & \makecell{Minimal lag \\ time of 5 days \\ (in manuscript)} &  \makecell{Minimal lag \\ time of 6 days} \\
\cmidrule{1-4}
Baden-W{\"u}rttemberg & 6 & 8 & 8\\
Bavaria & 12 & 8 & 12 \\
Berlin & 5 & - & - \\
Hesse & 6 & 7 & 7 \\
Lower Saxony & 6 & 7 & 7\\
North Rhine-Westphalia & 6 & 6 & 7 \\
Rhineland-Palatinate& 5 & 7 & 7\\
\bottomrule
\end{tabular}
\caption{\color{Gray}
Lag times (in days) for different German states using various minimal lag times for the ABC fit and GP training period. The standard choice employed in the findings of the manuscript is 5 days.\label{tab: sensitivity}}
\end{table}

In general, we note that a shorter minimal lag time gives a faster response in the data, and consequently a shorter lag time. This is not surprising, as we allow for a greater attribution of effects to school closures which may be the result of earlier interventions. Reassuringly, we find that extending the minimal lag time by a day, thereby further limiting the attribution of any response to school closures, does not significantly change the lag time. This alleviates some concern that 5 days, being solely the mean of a distribution of incubation times, is insufficient for the method to adequately capture the pre-intervention trajectory.
Expanding the sensitivity analysis to even longer minimal lag times is not meaningful: most lag times in the manuscript are 7 days, and by training the GP up until this point, any resulting lag times would likely be forced to be even longer. This amounts to forcing a signal, rather than searching for possible variability in the signal within a reasonably expected range. The effect of this can be seen in North Rhine-Westphalia.\newline

In Berlin, we find that a shorter minimal lag time of 4 days does allow for a deviation to be seen already on the fifth day when comparing the projected and observed trajectories. Inspection of Figure \ref{DE_BE} shows that there is a very gradual approach to the peak, which explains why more data points' signal needs to be attributed to school closures before a lag time can be determined.\newline
We encounter a similar limitation with Bavaria. In Figure \ref{DE_BY}, the data points from the 22\textsuperscript{nd} to at least the 25\textsuperscript{th} of March are very close to the boundary of the 75\% credible intervals of the GP. Although a visual inspection confirms that an 8 day lag time is consistent with a change in the data, small changes to the minimal lag can bring those observed points to fall just inside the boundary. When a 12 day lag time is analysed for Bavaria the relative reduction in the growth rate is 72\%, which is consistent with the other German states. The effect of interventions is consistently impacted in the state: either a smaller reduction the growth rate is seen compared to other states, or a similar reduction is obtained, but after a longer duration of time. 

\section{Testing data}
\subsection{Testing data for Germany}

\begin{figure}[H]
\centering
\includegraphics[width=130mm]{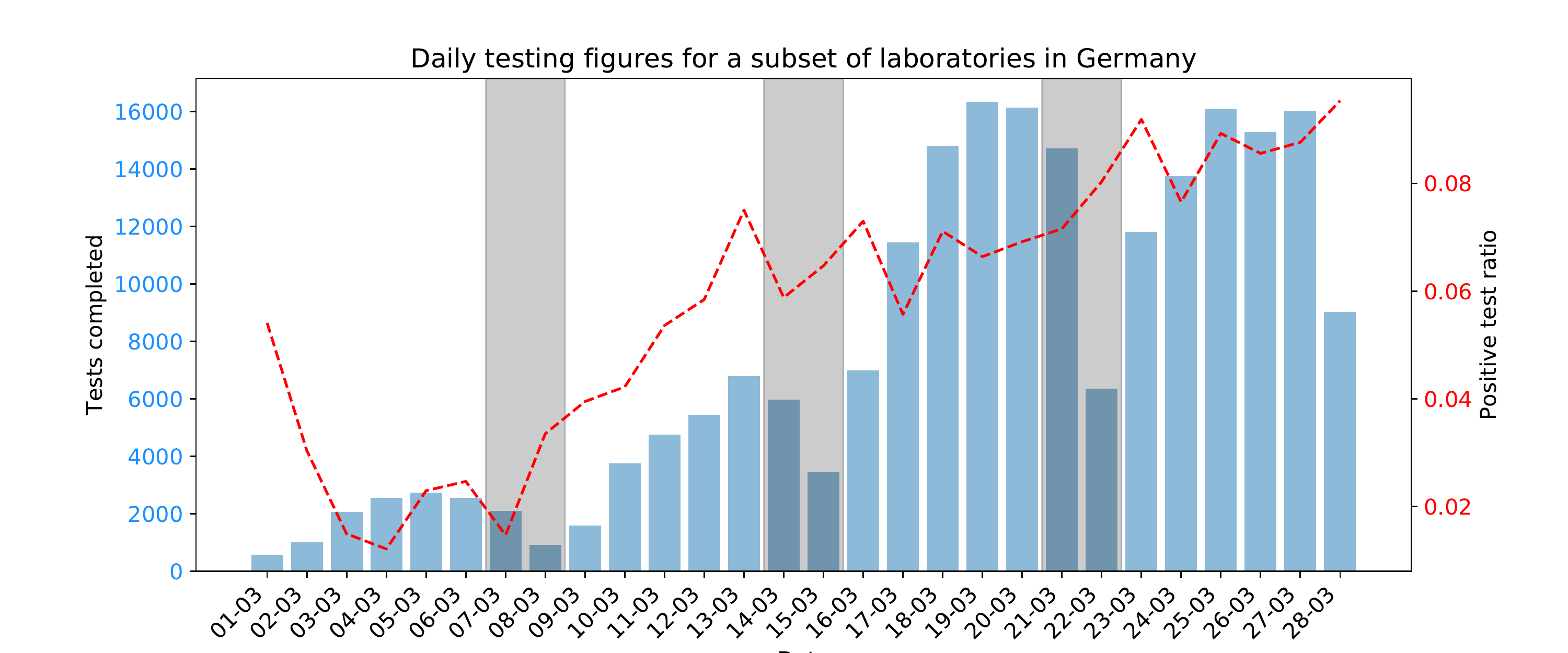}
\caption{\color{Gray}
Daily testing from a subset of German testing laboratories during March. Weekends are highlighted in grey. There is a periodic drop in testing occurring on weekends, particularly evident on Sundays. These drops do not coincide with any changes to the positive test ratio. }
\label{DE_daily_tests}
\end{figure}

Figure \ref{DE_daily_tests} shows the number of tests carried out per day in German medical laboratories, along with the positive test ratio over the same period. This is not equivalent to the total number of tests carried out in Germany, as not all laboratories provided this type of data; however, it can be used as an indication of general testing trends.
There is a weekend effect occurring in the testing data for Germany, with lower relative testing occurring on March 7\textsuperscript{th}-9\textsuperscript{th}, 14\textsuperscript{th}-16\textsuperscript{th} and 21\textsuperscript{st}-23\textsuperscript{rd}. No corresponding change is seen in the positive test ratio, indicating that case numbers were likely consistent across these periods. As such, any fall in confirmed cases over these periods can likely be attributed to reduced testing, rather than a response to intervention.
Ignoring the weekend effect, the number of tests carried out across the period between March 17\textsuperscript{th} and 27\textsuperscript{th} was fairly stable. As most school closures in Germany occurred on March 16\textsuperscript{th}, we can expect the confirmed cases over this period to provide a reasonable representation of the underlying epidemic.

\begin{figure}[H]
\centering
\includegraphics[width=130mm]{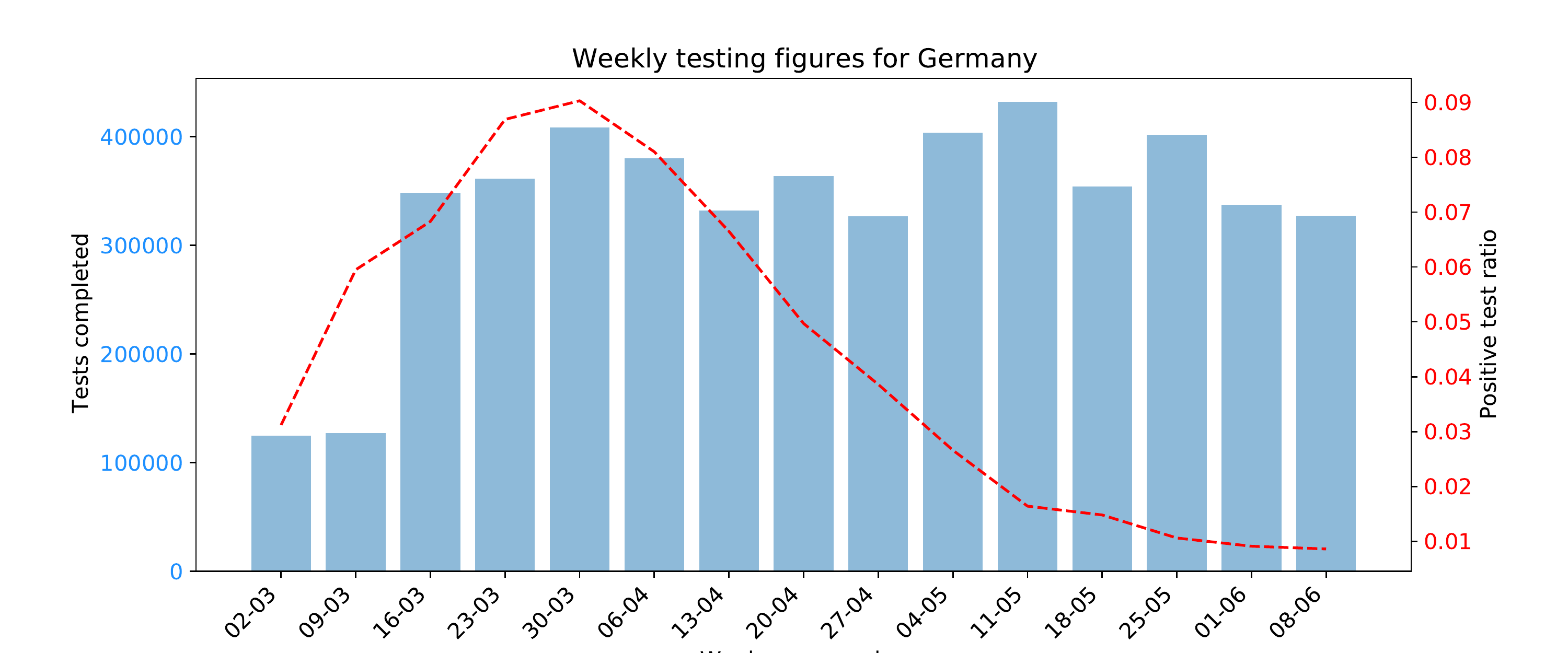}
\caption{\color{Gray}
Weekly testing in Germany remained consistent from March 18\textsuperscript{th}, however the weekend effect (see Figure \ref{DE_daily_tests}) was likely present across the entire period. There were no abrupt changes in the positive test ratio.}
\label{DE_weekly_tests}
\end{figure}

Daily testing data for Germany are not available after March. As such, it will be necessary to consider the weekly testing totals, which are made available through the RKI. Figure \ref{DE_weekly_tests} shows the weekly testing numbers for Germany, along with the weekly positive test ratio. Note that weekly testing data for Germany are released every Wednesday.

\subsection{Testing data for Scandinavia}
\begin{figure}[H]
\centering
\includegraphics[width=130mm]{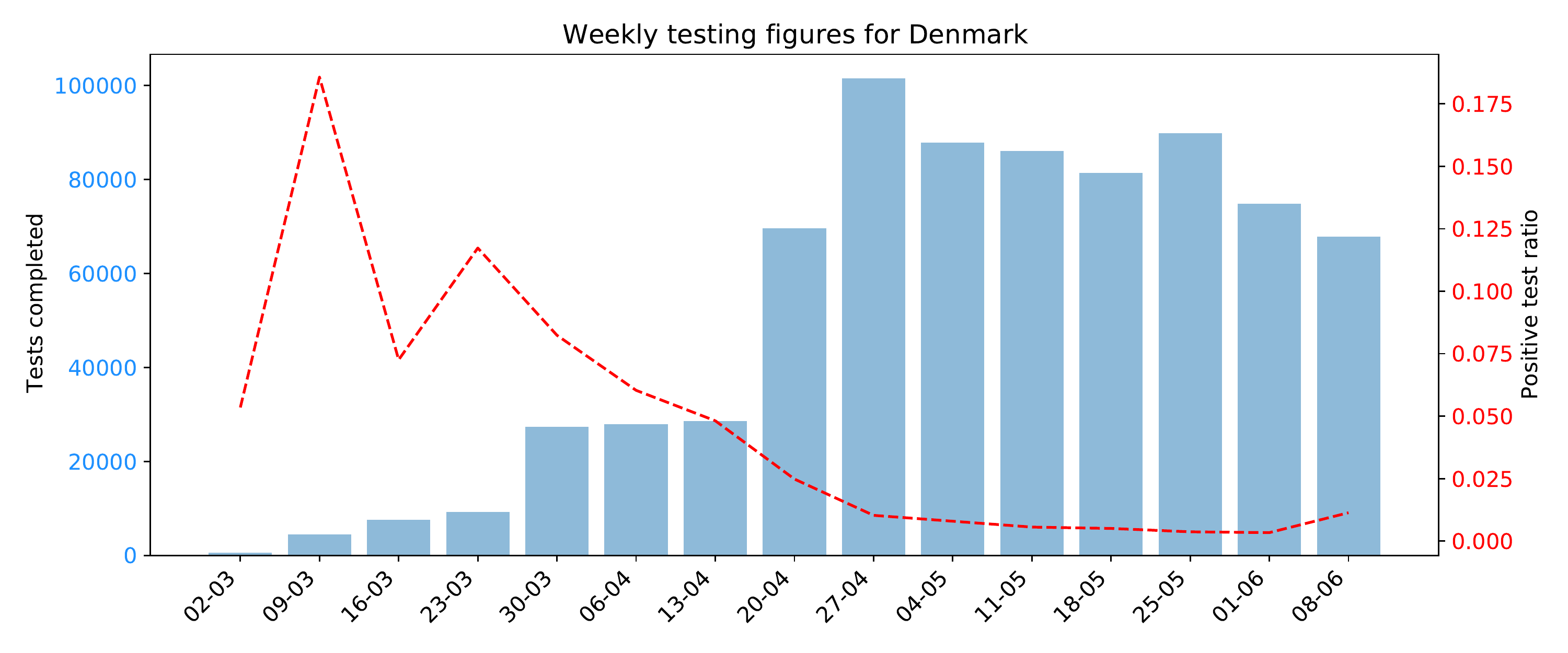}
\caption{\color{Gray}
Weekly testing in Denmark was not consistent across the period of this investigation, and so confirmed cases up to April 20\textsuperscript{th} cannot be relied upon to provide a reliable representation of the underlying epidemic.}
\label{DK_weekly_tests}
\end{figure}

\begin{figure}[H]
\centering
\includegraphics[width=130mm]{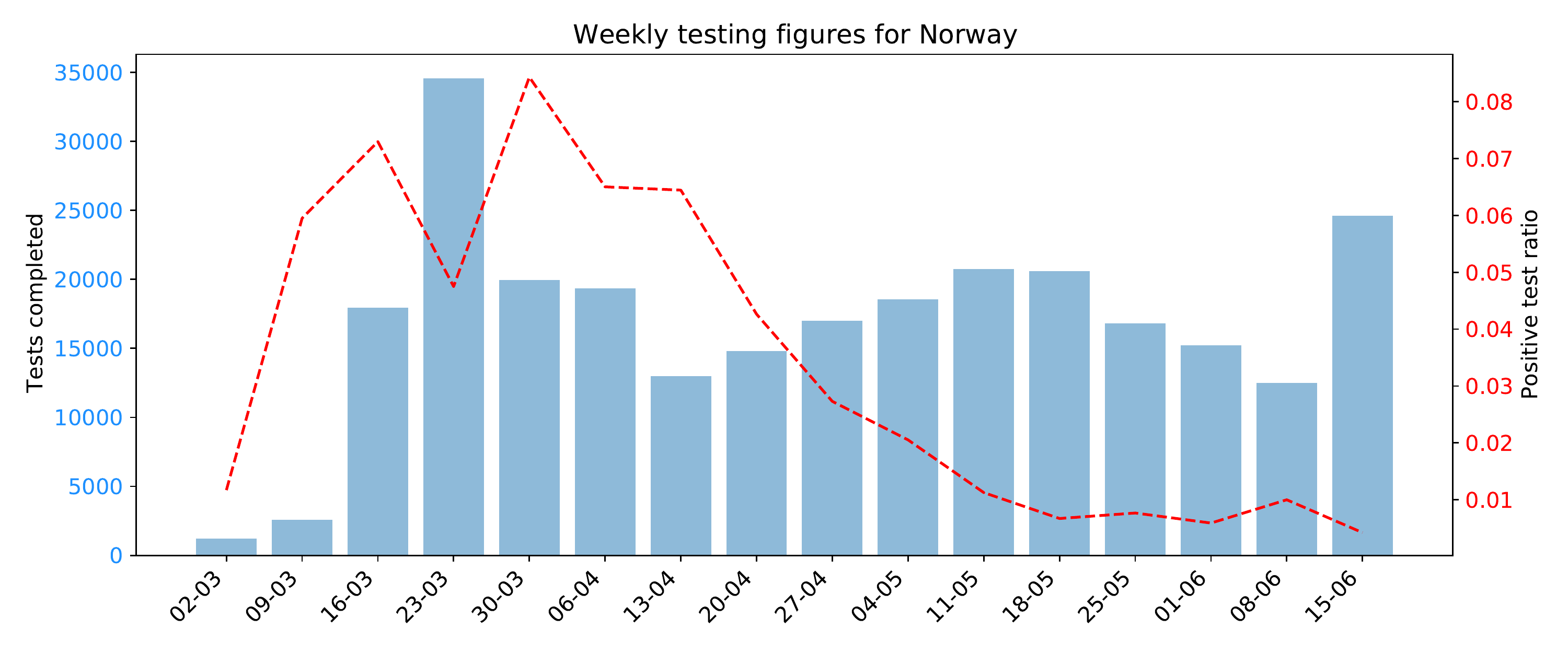}
\caption{\color{Gray}
Norway saw inconsistent testing during March, making confirmed cases an inappropriate metric for assessing school closures. More consistent testing was apparent in April and May.}
\label{NO_weekly_tests}
\end{figure}

Both Denmark and Norway saw a similar weekend effect in testing numbers, with midweek testing figures roughly 50\% higher than weekend figures in Denmark, and almost three times higher in Norway. The weekly testing figures for Denmark and Norway are shown in Figure \ref{DK_weekly_tests} and \ref{NO_weekly_tests} respectively.\newline
Denmark displays two clear increases in testing capacity between March 23\textsuperscript{rd} and 30\textsuperscript{th} and again between April 13\textsuperscript{th} and 20\textsuperscript{th}. The increase in late March, combined with a relatively high positive test ratio,  indicates that confirmed cases during this period might not be a suitable metric.\newline 
Similarly in Norway there was a large increase in testing in the week commencing March 16\textsuperscript{th}, very close to the date of school closures. As such, for both Norway and Denmark it will be necessary to consider hospitalisations as a metric for assessing the dynamics of the epidemic. 

\begin{figure}[H]
\centering
\includegraphics[width=130mm]{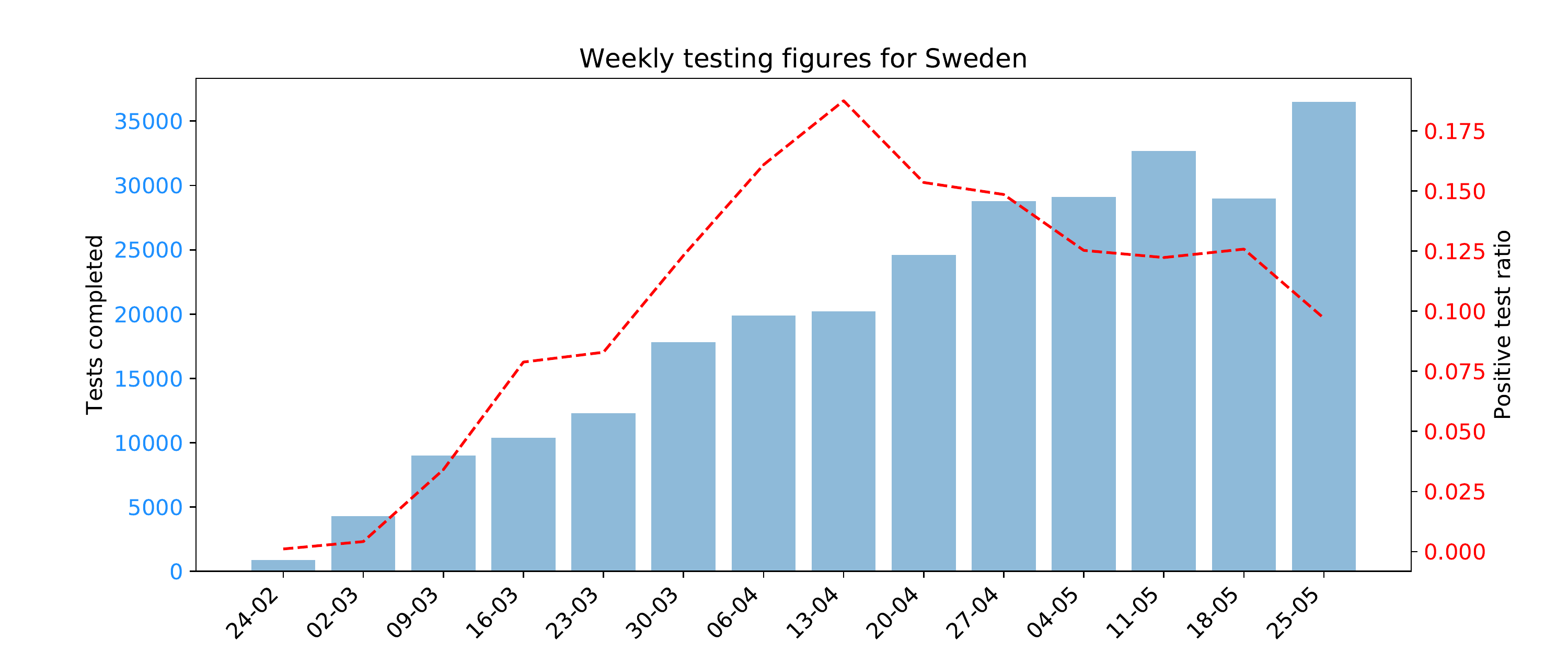}
\caption{\color{Gray}
Reported weekly tests carried out in Sweden. These numbers rose throughout March and April.}
\label{SE_weekly_tests}
\end{figure}

The weekly testing figures for Sweden are highlighted in Figure \ref{SE_weekly_tests}, along with the positive test ratio for the same period. Testing rates around the time of school closures (March 18\textsuperscript{th}) were generally increasing, with a large increase occurring during the week beginning March 30\textsuperscript{th}. This increase was accompanied by an increase in positive test ratio, indicating an increasing capability to identify and test infected individuals, or a change in testing policy. As a result, it will not be possible to attribute any change in the increase in case numbers after March 30\textsuperscript{th} solely to the effect of interventions.

Despite the changes in testing rates around the time of school closures, Sweden is still able to provide very useful insight. The decision by the Swedish government to (a) leave schools open to all students under the age of 16 and (b) do so with a background of limited social interventions is useful for partially decoupling the effect of school closures from other controls.

\section{School closure analyses}
We present the equivalent of Table \ref{tab1} in the manuscript, but expressed via the doubling time ($\ln(2)/\textrm{growth rate}$).

\begin{table}[H]
\ra{1.3}
\centering
\begin{tabular}{@{}lcccc@{}}
\toprule
\textbf{State} & \makecell{Lag time \\ (days)} & \makecell{Pre-response \\ doubling time \\ (days)} &  \makecell{Modelled \\ post-response \\ doubling time (days)} & \makecell{Observed \\ post-response \\doubling time (days)}\\
\cmidrule{1-5}
Baden-W{\"u}rttemberg & 8 &\makecell{4.9 \\(3.8 - 6.6)}& \makecell{4.2 \\(3.7 - 4.7)}& \makecell{13.5 \\(7.7 - 51.9)}\\ 
Bavaria & 8 &\makecell{3.2 \\(2.7 - 3.9)}& \makecell{3.2 \\(3.1 - 3.3)}& \makecell{6.3 \\(4.7 - 9.6)} \\ 
Berlin & --$^*$ &\makecell{4.8 \\(3.7 - 6.7)}& -- & -- \\ 
Hesse & 7 &\makecell{2.8 \\(2.3 - 3.6)}& \makecell{2.5 \\(2.4 - 2.6)}& \makecell{10.3 \\(5.9 - 40.0)} \\ 
Lower Saxony & 7 &\makecell{3.1 \\(2.6 - 3.9)}& \makecell{3.0 \\(2.8 - 3.3)}& \makecell{10.0 \\(6.5 - 21.9)} \\ 
North Rhine-Westphalia & 6 &\makecell{3.6 \\(3.0 - 4.4)}& \makecell{3.4 \\(3.3 - 3.5)}& \makecell{11.3 \\(7.2 - 26.3)}  \\ 
Rhineland-Palatinate& 7 &\makecell{2.7 \\(2.2 - 3.4)}& \makecell{3.0 \\(2.7 - 3.3)}& \makecell{16.0 \\(8.1 - 1188.4)} \\
\bottomrule
\end{tabular}
\caption{\color{Gray}
Comparison of estimated lag time and pre- and post-intervention doubling times in different German states.\newline
$^*$The peak in daily incidence is reached before a response is seen in the data. A lag time which may be attributable to school closures therefore cannot be determined.}
\end{table}

\begin{figure}[H]
\centering
\includegraphics[width=130mm]{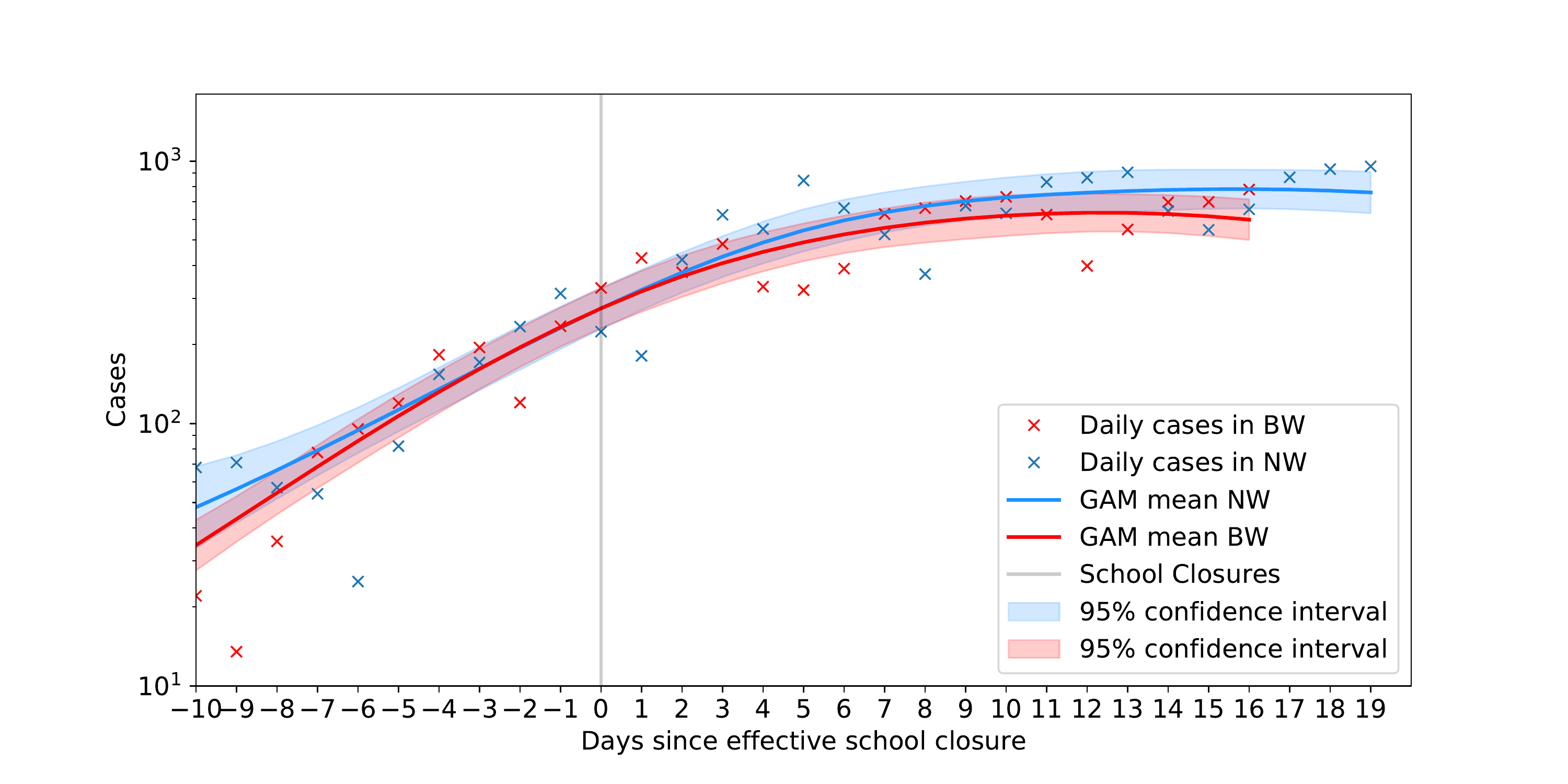}
\caption{\color{Gray} Daily cases for Baden-W\"{u}rttemberg and North Rhine-Westphalia when corrected for the three-day shift in school closure between the two. The effective day of school closure in both states is shown by the solid grey line. The smoothed trajectory, obtained via a GAM analysis, is shown by the solid lines with the shaded regions representing the 95\% confidence intervals. The data for  North Rhine-Westphalia are scaled so that the GAM smoothed incidence are identical on the day of closures.\newline
There is very good agreement between the two data streams despite the time difference in the school closure, suggesting comparable underlying transmission in the two states following school closure. Additionally the lag times are comparable between the two states.
} 
\label{NW_BW_comp}
\end{figure}

In Figure \ref{NW_BW_comp} we compare states which closed schools at different times to examine the robustness of the response time being associated with school closures, and not a fixture particular to a certain day. 
The comparison between states with different dates of school closure is limited to Baden-W\"{u}rttemberg and North Rhine-Westphalia for a number of reasons. Firstly, under the hypothesis that the effectiveness of interventions, or the timing of a response, would depend on the community prevalence, we are limited to comparing Baden-W\"{u}rttemberg with Bavaria, and North Rhine-Westphalia. However, Bavaria imposed stay-at-home orders and held local elections on different dates than the rest of the analysed states. In order to fix as many other factors as possible to be constants, we are left only with the comparison between Baden-W\"{u}rttemberg and North Rhine-Westphalia.\newline
A similar comparison using Berlin, which also closed schools on the 17\textsuperscript{th} of March was not possible, due to the lack of a lag time resulting from the analysis. Furthermore, Berlin had a different timeline of interventions than other states, making for an unjust comparison.\newline

We shift the cases in Baden-W\"{u}rttemberg back in time by three days, so as to coincide the dates of effective school closure (March 14\textsuperscript{th} and 17\textsuperscript{th}). Further, to aid comparison we rescale the cases in Baden-W\"{u}rttemberg by a multiplicative constant (approximately 0.76), so that the cases in both states are identical on the day of school closure. We note that the pure exponential growth rate is unchanged by these transformations. The profiles of the data can now be compared.

The two states appear to be comparable both in terms of the overall cases following school closure, as well as the time taken until a response from an intervention can be observed in the data. While the lag times are not identical, sensitivity analysis (see Table \ref{tab: sensitivity}) suggests that these lag times overlap. Furthermore, it is reassuring that there does not seem to be a fixed date following school closures when a response is seen. Were this the case, we would expect Baden-W\"{u}rttemberg to have a lag time which was 3 days shorter than North Rhine-Westphalia.
This suggests a signal in both data sets following school closure. Clearly it would be unrealistic to assume school closures to be wholly responsible for the observed fall in cases, but the above-detailed correlations suggest that they may have partially contributed to the overall effect.

\subsubsection*{Bavaria}
\begin{figure}[H]
\centering
\includegraphics[width=130mm]{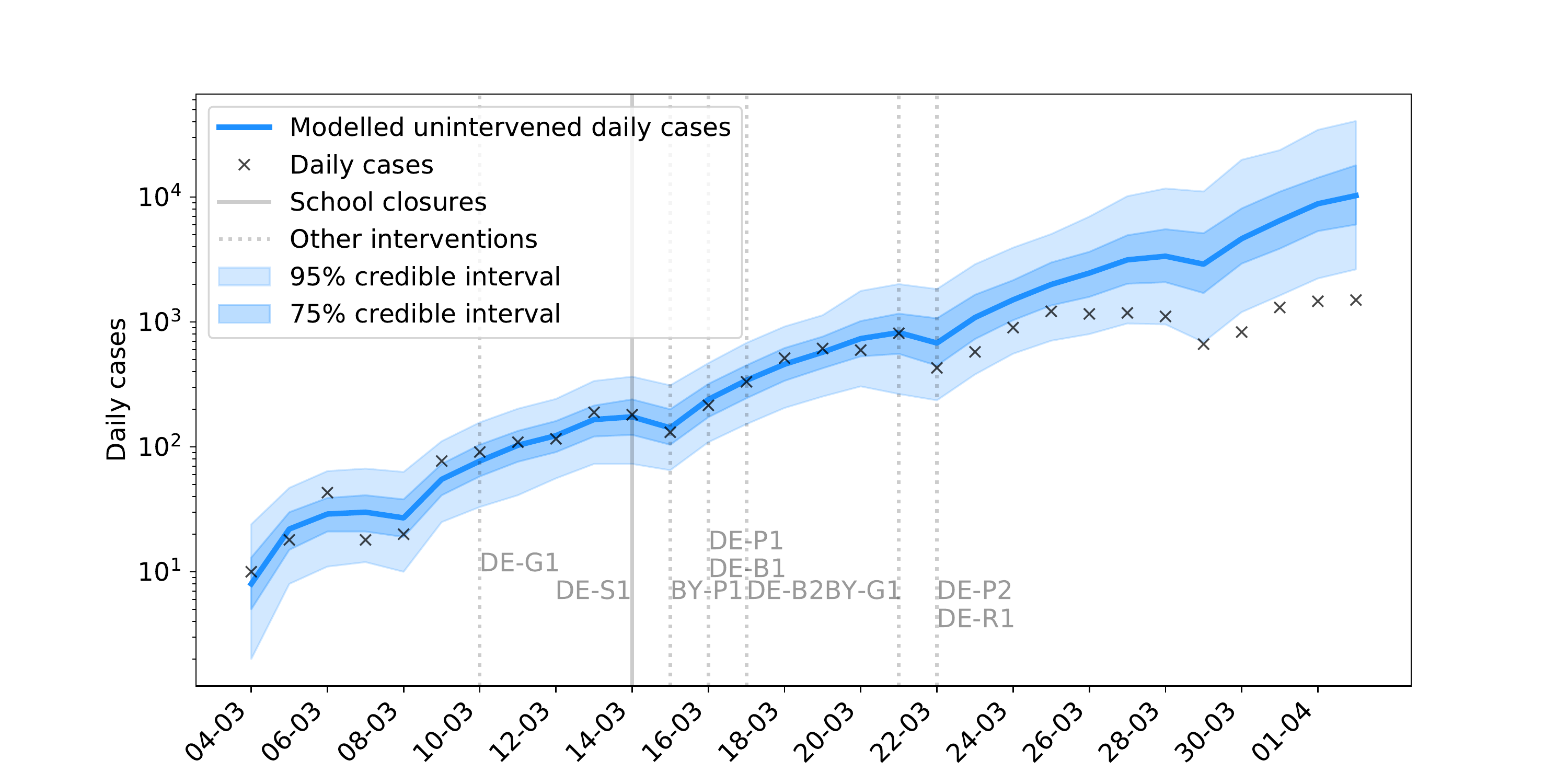}
\caption{\color{Gray}Modelled and observed cases in Bavaria.} 
\label{DE_BY}
\end{figure}

Bavaria saw a sustained decrease in epidemic growth rate occurs 8 days after school closures (Figure \ref{DE_BY}).
Bavaria saw the following interventions around the time of school closures:

\begin{itemize}
    \item 10/03 - Banned gatherings of more than 1000 people (DE-G1).
    \item 14/03 - \textbf{School closures} (effective date, DE-S1).
    \item 15/03 - Local elections went ahead, with a high turnout (BY-P1). A large number of votes were submitted by post.
    \item 16/03 - State of emergency declared; shut borders with France (FR), Switzerland (CH), Austria (AT), Denmark (DK) and Luxembourg (LU) (DE-B1); closure of non-essential businesses and public services (DE-P1).
    \item 17/03 - Shut borders with EU (DE-B2).
    \item 21/03 - Stay at home orders, with exceptions for essential trips, and banned gatherings of more than 2 people (BY-G1);
    \item 22/03 - Banned all social events and gatherings (DE-P2); closure of non-essential retail and leisure, with exceptions for restaurant takeout (DE-R1).
\end{itemize}

\subsubsection*{Berlin}
\begin{figure}[H]
\centering
\includegraphics[width=130mm]{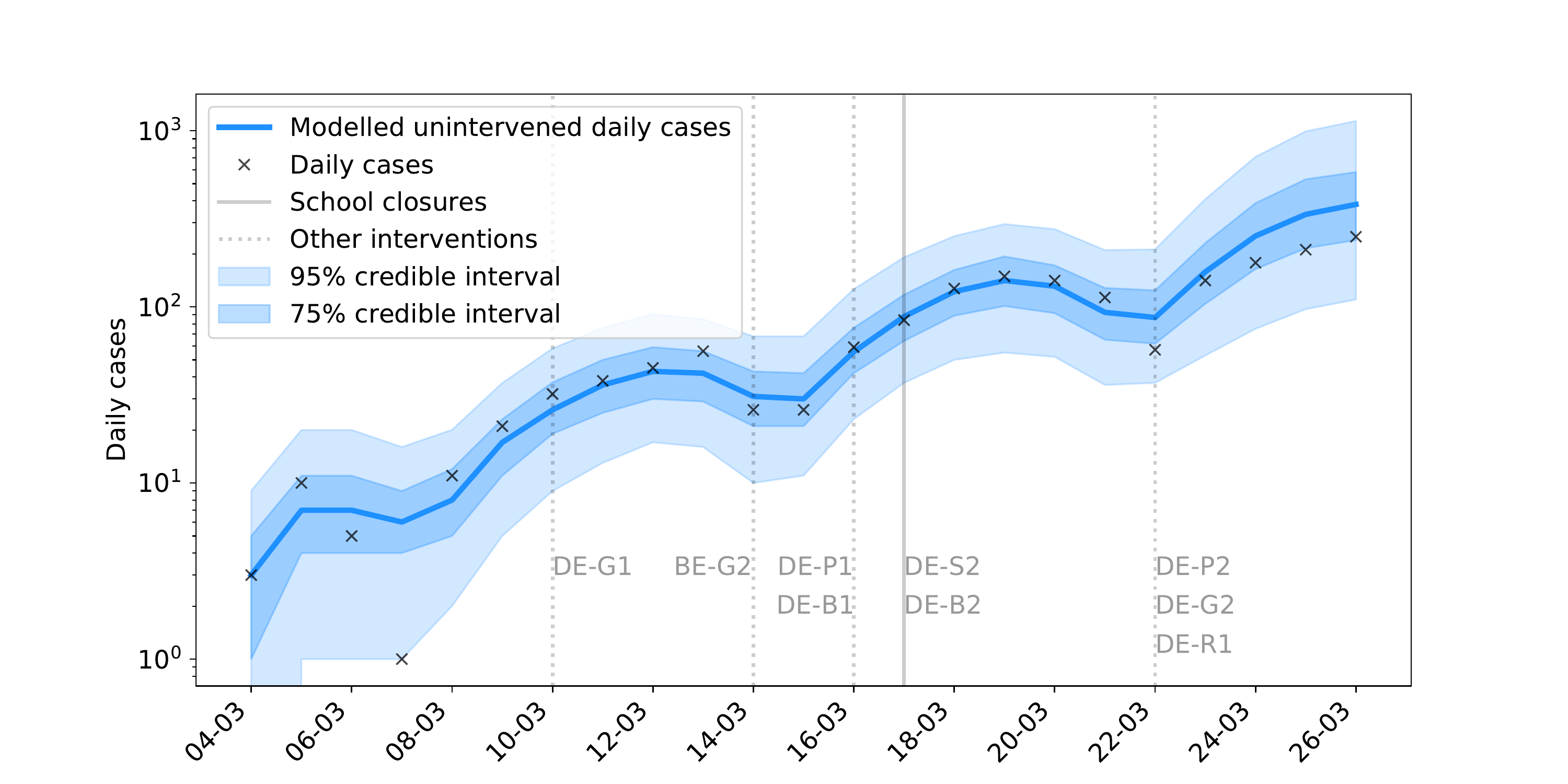}
\caption{\color{Gray}Modelled and observed cases in Berlin.} 
\label{DE_BE}
\end{figure}

Berlin did not see a response which may be partially attributable to school closures, as a peak in daily cases was reached before a signal was observed in the data (Figure \ref{DE_BE}). This suggests that the subsequent decline in cases in the state was brought about by other interventions. Testing rates throughout this period were consistent, indicating that confirmed cases are likely a good indicator of general trends in the epidemic (Figure \ref{DE_BE}).

Berlin saw the following interventions around the time of school closures:

\begin{itemize}
    \item 10/03 - Banned gatherings of more than 1000 people (DE-G1).
    \item 14/03 - Banned events with more than 50 people (BE-G1).
    \item 16/03 - Shut borders with France (FR), Switzerland (CH), Austria (AT), Denmark (DK) and Luxembourg (LU) (DE-B1); closure of non-essential business and public service (DE-P1).
    \item 17/03 - \textbf{School closures} (DE-S2); shut borders with EU (DE-B2).
    \item 22/03 - National stay at home orders, with exceptions for essential trips, and banned gatherings of more than 2 people (DE-G2); banned all social events and gatherings (DE-P2); closure of non-essential retail and leisure, with exceptions for restaurant takeout (DE-R1).
\end{itemize}

\subsubsection*{North Rhine-Westphalia}
\begin{figure}[H]
\centering
\includegraphics[width=130mm]{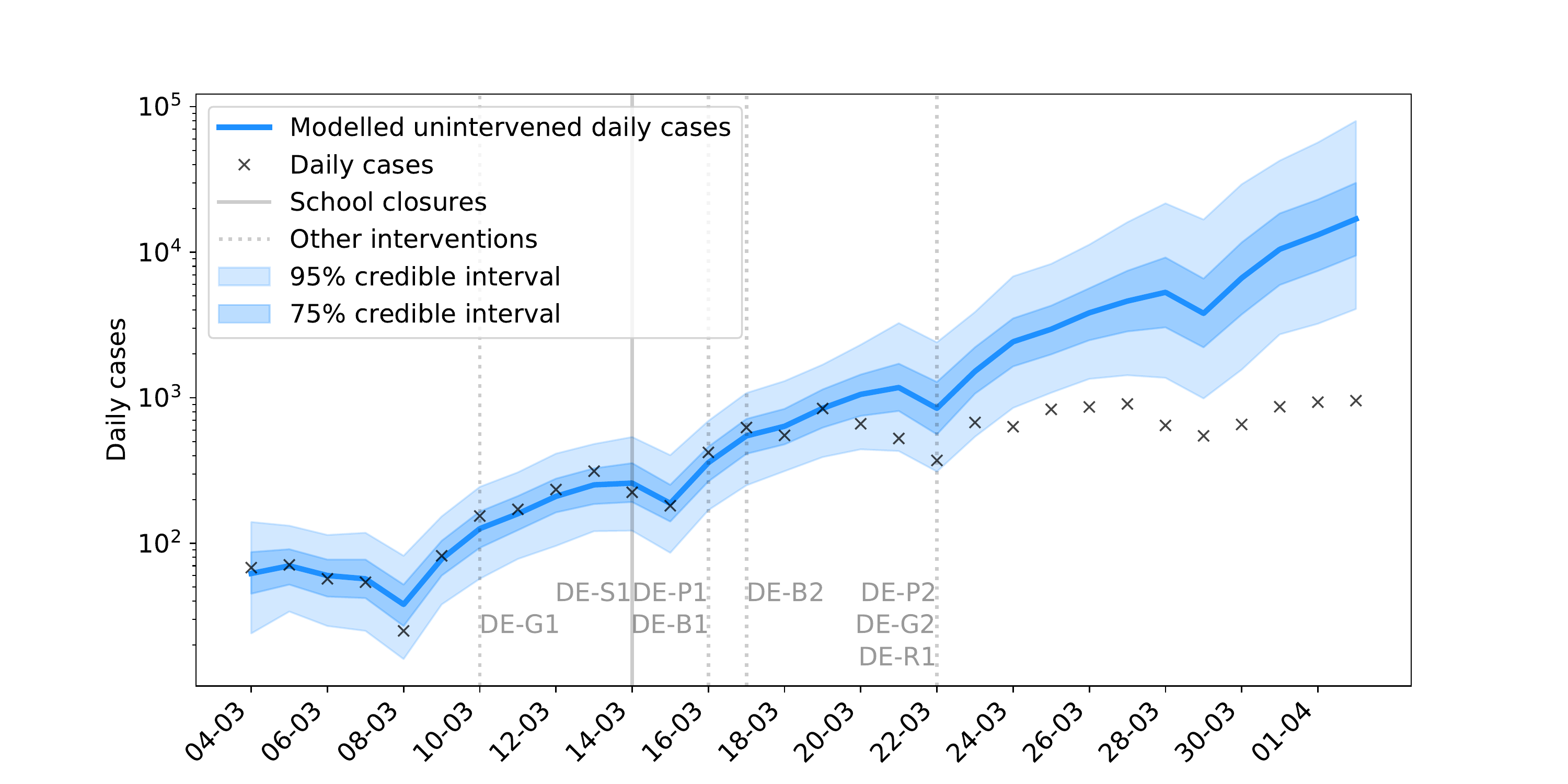}
\caption{\color{Gray}Modelled and observed cases in North Rhine-Westphalia.} 
\label{DE_NW}
\end{figure}

North Rhine-Westphalia saw a sustained drop in epidemic growth rate occurs on the 20\textsuperscript{th}, 6 days after school closures (Figure \ref{DE_NW}).

\begin{itemize}
    \item 05/03 - \textbf{Local school closures in Heinsberg.}
    \item 10/03 - Banned gatherings of more than 1000 people (DE-G1).
    \item 14/03 - \textbf{School closures} (effective date, DE-S1).
    \item 16/03 - Shut borders with France (FR), Switzerland (CH), Austria (AT), Denmark (DK) and Luxembourg (LU) (DE-B1); closure of non-essential business and public service (DE-P1).
    \item 17/03 - Shut borders with EU (DE-B2).
    \item 22/03 - National stay at home orders, with exceptions for essential trips, and banned gatherings of more than 2 people (DE-G2); banned all social events and gatherings (DE-P2); closure of non-essential retail and leisure, with exceptions for restaurant takeout (DE-R1).
\end{itemize}

\subsubsection*{Denmark}
\begin{figure}[H]
\centering
\includegraphics[width=130mm]{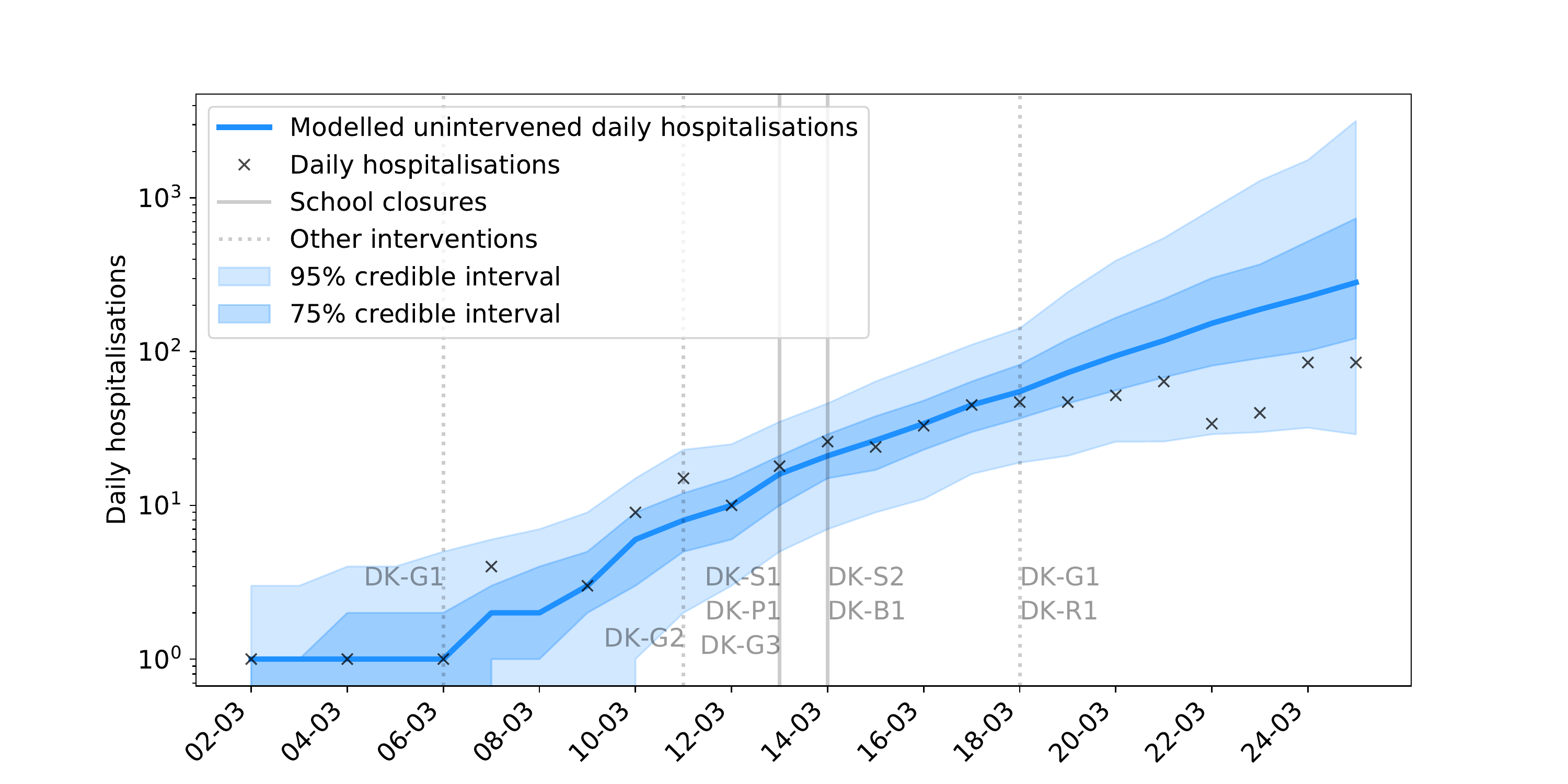}
\caption{\color{Gray}Modelled and observed daily hospitalisations in Denmark. The peak in daily hospitalisations occurs too soon after school closures to suggest that those interventions could have contributed to the observed reduction in the growth rate of cases.} 
\label{DK_close_i}
\end{figure}

Denmark saw a staged closing of schools, with primary school closing on Friday 13\textsuperscript{th} of March, and all other schools following on Monday the 16\textsuperscript{th} of March. The effective date of secondary school closures is taken to be Saturday 14\textsuperscript{th} of March.\newline
While there is a decrease in growth rate in the period following school closures (Figure \ref{DK_close_i}), it is not attributable to school closures, as the peak in hospitalisations occurs much sooner than the expected time from infection to hospitalisation (10-14 days).

Denmark saw the following interventions introduced around the same time as school closures:
\begin{itemize}
    \item 06/03 - Recommendation against events with more than 1000 people (DK-G1).
    \item 11/03 - Recommendation against public gatherings of more than 100 people (DK-G2).
    \item 13/03 - \textbf{School closure for students aged 16 or over} (DK-S1); closure of non-essential businesses, public sports or leisure facilities, and cultural institutions (DK-P1); work from home order for non-essential workers, with exceptions for essential travel (DK-G3).
    \item 14/03 - \textbf{School closures} (effective date, DK-S2); borders closed (DK-B1).
    \item 18/03 - Banned gatherings of more than 10 people (DK-G4); closed non-essential retail, with exceptions for takeaway and food delivery (DK-R1).
\end{itemize}

\subsubsection*{Norway}
Norway closed schools at the same time as introducing a range of other restrictions on social life. As such it is not possible to attribute the observed change in hospital admissions solely to school closures. It is notable, however, that the observed reduction in hospitalisations is comparable in Denmark and Norway; both countries which simultaneously targeted schools and non-essential businesses.

\begin{figure}[H]
\centering
\includegraphics[width=130mm]{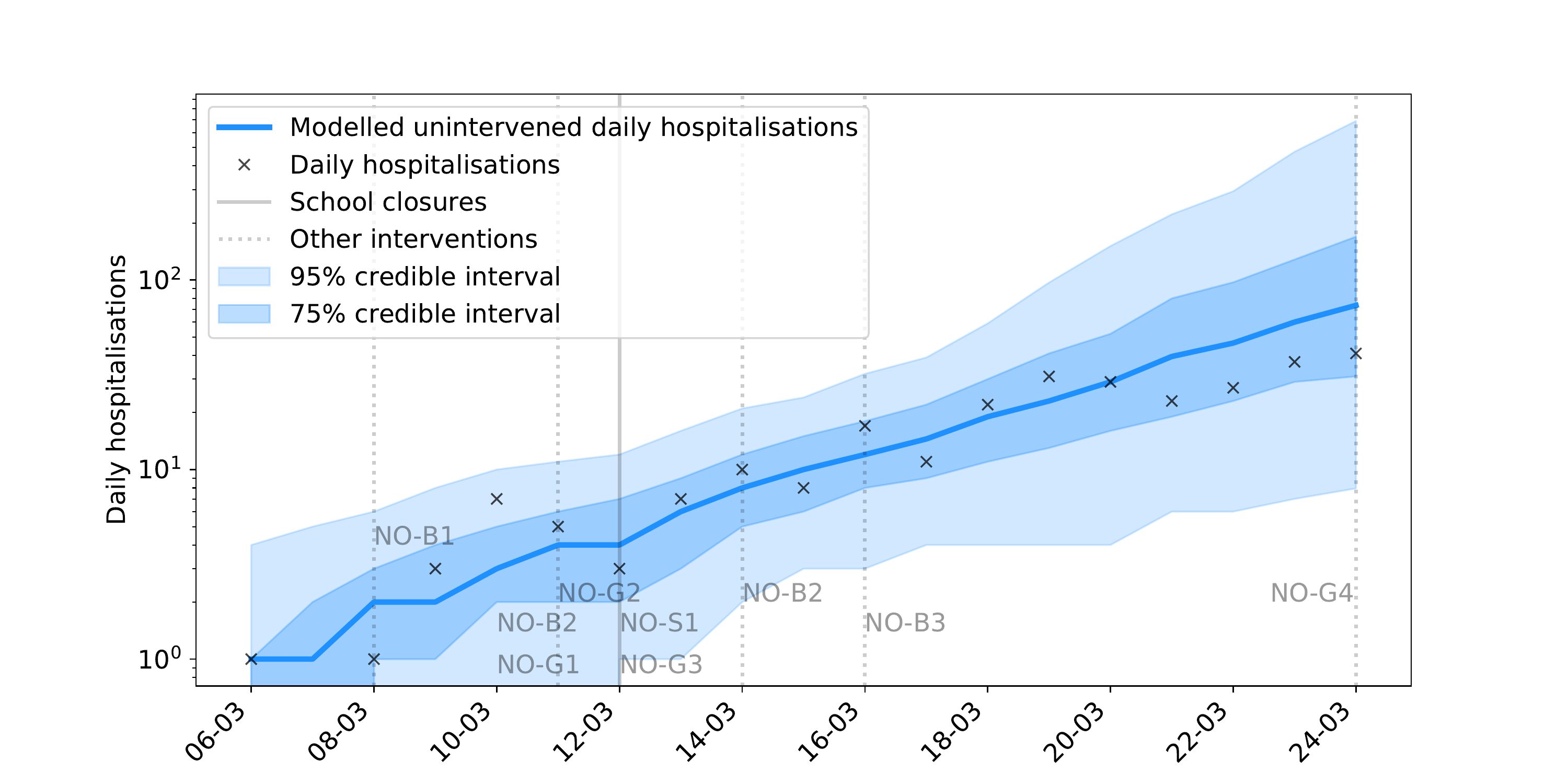}
\caption{\color{Gray}Modelled and observed hospitalisations in Norway.} 
\label{NO_close_i}
\end{figure}

For completeness, we fit the GP model to daily hospitalisations (Figure \ref{NO_close_i}). However the short training period for the model resulted in a wide credible interval for this projection. Furthermore, as with Denmark, the proximity of school closures to the point of peak incidence makes it difficult to adequately assess either the lag time or the change in growth rate occurring after closures.  As such, it is difficult to draw any firm conclusions from the data.

Norway saw the following interventions introduced around the same time as school closures:
\begin{itemize}
    \item 08/03 - Docking restrictions for large ships (NO-B1).
    \item 10/03 - Flight restrictions (NO-B2); working from home encouraged (NO-G1).
    \item 11/03 - Recommendation against gatherings with more than 500 people (NO-G2).
    \item 12/03 - \textbf{School closures} (NO-S1); closure of non-essential businesses, public leisure facilities, and non-essential retail; essential travel only (NO-G3). Exceptions were made for venues serving food, where social distancing could be observed.
    \item 14/03 - Recommendations against foreign travel (NO-B2).
    \item 16/03 - Border and entry restrictions (NO-B3).
    \item 24/03 - Banned outdoor gatherings of more than 5 people (NO-G4).
\end{itemize}

\section{School reopening analyses}
We here present results in support of the main results of the paper, but which are not essential to the exposition of our findings.
\begin{figure}[H]
\centering
\includegraphics[width=130mm]{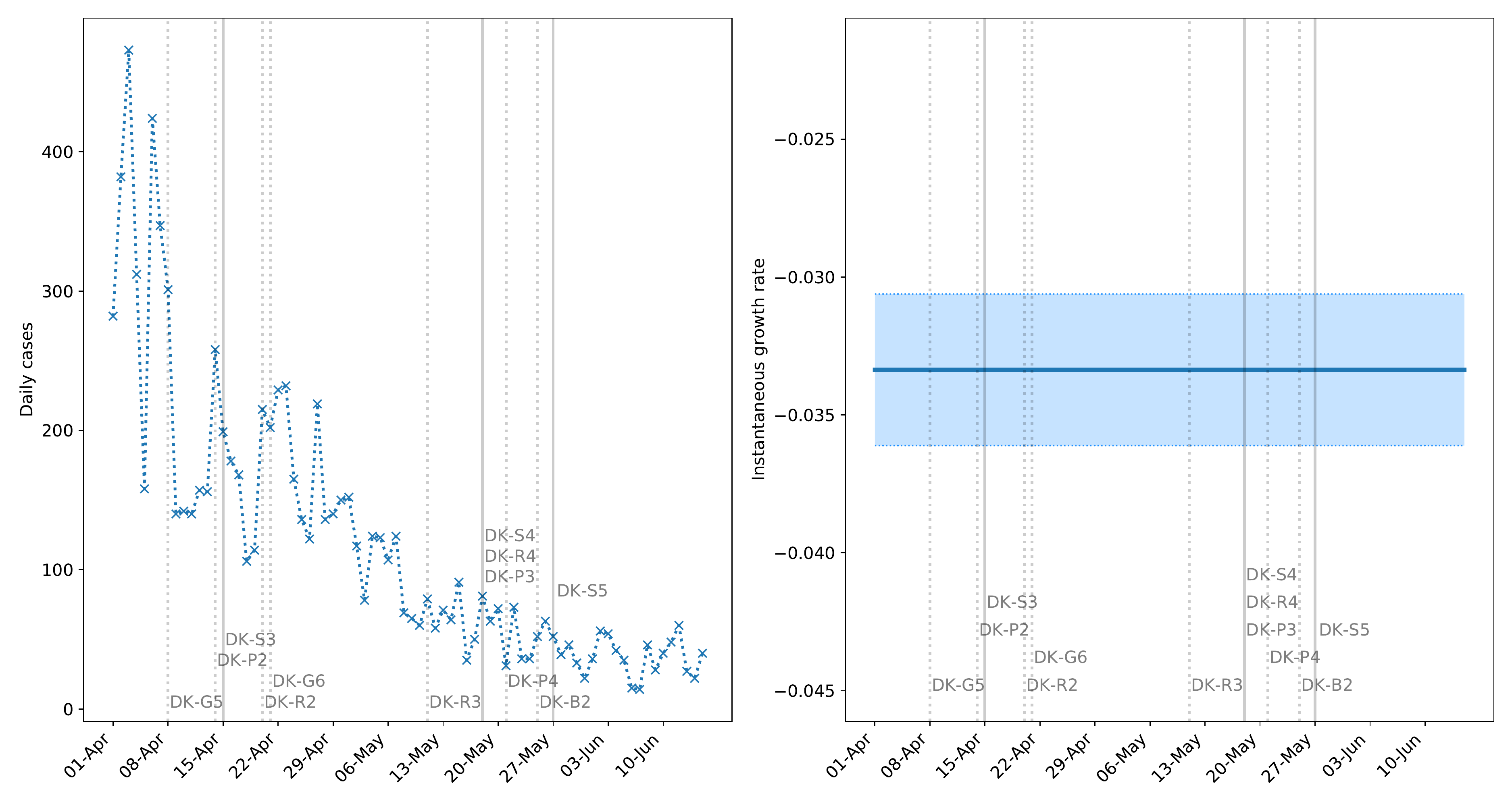}
\caption{\color{Gray}
Reported daily confirmed cases in Denmark. The left panel shows new cases, and the right panel shows the instantaneous growth rate (shaded regions are 95\% confidence intervals). Solid vertical lines indicate when students returned to school, and dashed lines indicate other loosened measures. We present these numbers in support of the observations made for daily hospital admissions due to the larger numbers of cases recorded here. These results are not qualitatively different from those obtained from hospitalisation data, but support the conclusions which are harder to draw from that data set due to the longer delay from infection to hospitalisation.\newline
We refrain from interpreting positive cases before early May due the increase in testing from the 20\textsuperscript{th} of April.} 
\label{DK-cases}
\end{figure}
Sentinel survey information indicates the following for staff in different educational settings. A smaller proportion of staff working with young children have tested positive compared to staff working with older students. However, these numbers alone do not distinguish between infection acquired from the students, and infection acquired elsewhere (Table \ref{DK-sentinel}).
\begin{table}[H]
\ra{1.3}
\centering
\begin{tabular}{@{}lccc@{}}
\toprule
\textbf{Educational Level} & \textbf{Tested Population (\%)} & \textbf{Positive Tests (\%)} & \textbf{Tests}\\
\cmidrule{1-4}
Nursery & 9.63 &1.18& 593\\ 
Kindergarten & 12.85 &0.90& 2773\\ 
Primary school (ages 7 to 15/16) & 13.36 &1.23& 14855\\ 
Secondary school (ages 16 to 19) & 8.79 &1.35& 3343\\ 
Higher education & 8.30 &1.85& 3354\\ 
Adult education & 13.22 &1.43& 2875\\
\bottomrule
\end{tabular}

\caption{\color{Gray}
A comparison of tests carried out among staff working in different stages of the Danish educational and childcare sector dated June 2\textsuperscript{nd}. We indicate the proportion of tested staff relative to estimated employee numbers in each group, and the percentage of those tested who test positive. For reference, the absolute numbers of tests are also shown.}
\label{DK-sentinel}
\end{table}

\end{document}